\documentclass[ejs]{imsart}

\RequirePackage[OT1]{fontenc}
\RequirePackage{amsthm,amsmath,mathabx}
\RequirePackage[numbers]{natbib}
\RequirePackage[colorlinks,citecolor=blue,urlcolor=blue]{hyperref}
\usepackage{times,epsfig,graphics,latexsym}

\pubyear{0000}
\volume{0}
\issue{0}
\firstpage{0}
\lastpage{0}

\def\wt{\widetilde}

\def\wh{\widehat}

\def\log{\hbox{log}}

\def\boxit#1{\vbox{\hrule\hbox{\vrule\kern6pt
          \vbox{\kern6pt#1\kern6pt}\kern6pt\vrule}\hrule}}

\def\bse{\begin{eqnarray*}}
\def\ese{\end{eqnarray*}}
\def\be{\begin{eqnarray}}
\def\ee{\end{eqnarray}}
\def\bq{\begin{equation}}
\def\eq{\end{equation}}

\def\wh{\widehat}

\def\trans{^{\rm T}}

\newtheorem{lemma}{Lemma}

\def\bft{{\bf t}}

\def\bfx{{\bf x}}

\def\bbeta{{\mbox{\boldmath $\beta$}}}

\def\bdelta{{\mbox{\boldmath $\delta$}}}

\def\bgamma{{\mbox{\boldmath $\gamma$}}}

\def\bomega{{\mbox{\boldmath $\omega$}}}

\def\bSigma{{\mbox{\boldmath $\Sigma$}}}

\def\btheta{{\mbox{\boldmath $\theta$}}}

\def\delete#1{\iffalse #1 \fi}

\def\argmin{\mathop{\rm argmin}}

\def\pmb#1{\setbox0=\hbox{#1}%
    \kern-.025em\copy0\kern-\wd0
    \kern.05em\copy0\kern-\wd0
    \kern-.025em\raise.0433em\box0 }
\def\pmbh#1#2{\setbox0=\hbox{#1}%
    \setbox1=\hbox{#2}%
    \kern-.025em\copy0\kern-\wd0
    \kern.05em\copy1\kern-\wd0
    \kern-.025em\raise.0433em\box0 }

\def\binom#1#2{{#1\choose #2}}
\def\frac#1#2{{#1\over#2}}

\def\boxit#1{\vbox{\hrule\hbox{\vrule\kern6pt
   \vbox{\kern6pt#1\kern6pt}\kern6pt\vrule}\hrule}}

\def\listing#1{\vskip 4mm\begin{verbatim}\input#1 \vskip 4mm}
\def\thick#1{\hbox{\rlap{$#1$}\kern0.25pt\rlap{$#1$}\kern0.25pt$#1$}}

\def\wt{\widetilde}
\def\wh{\widehat}

\def\bft{{\bf t}}

\def\bfx{{\bf x}}

\def\bfzero{{\bf 0}}

   \def\boldA{{\mbox{\boldmath $A$}}}
\def\boldb{{\mbox{\boldmath $b$}}}   \def\boldB{{\mbox{\boldmath $B$}}}
   \def\boldC{{\mbox{\boldmath $C$}}}
   \def\boldD{{\mbox{\boldmath $D$}}}
\def\bolde{{\mbox{\boldmath $e$}}}

\def\boldq{{\mbox{\boldmath $q$}}}   \def\boldQ{{\mbox{\boldmath $Q$}}}

   \def\boldT{{\mbox{\boldmath $T$}}}
\def\boldu{{\mbox{\boldmath $u$}}}   
\def\boldv{{\mbox{\boldmath $v$}}}   
   
   \def\boldX{{\mbox{\boldmath $X$}}}
   
\def\boldz{{\mbox{\boldmath $z$}}}   \def\boldZ{{\mbox{\boldmath $Z$}}}

\def\pmbh{{\pmb h}}

\def\calA{{\cal A}}

\def\calG{{\cal G}}
\def\calH{{\cal H}}

\def\calN{{\cal N}}

\def\calS{{\cal S}}

\def\stgbeta{{\thick{\scriptstyle{\beta}}}}	     
\def\stggamma{{\thick{\scriptstyle{\gamma}}}}

\startlocaldefs
\numberwithin{equation}{section}
\theoremstyle{plain}
\newtheorem{Theorem}{Theorem}[section]
\newtheorem{Cor}{Corralary}[section]
\endlocaldefs

\begin{document}

\begin{frontmatter}
\title{Additive Partially Linear Models for Massive Heterogeneous Data}
\runtitle{APLMs for Heterogeneous Data}

\begin{aug}

\author{\fnms{Binhuan} \snm{Wang} 
\ead[label=e1]{binhuan.wang@nyumc.org}}
\address{Department of Population Health\\
New York University School of Medicine\\
\printead{e1}}

\author{\fnms{Yixin} \snm{Fang} \thanksref{t1}
\ead[label=e2]{yixin.fang@njit.edu}}
\address{Department of Mathematical Sciences\\
New Jersey Institute of Technology\\
\printead{e2}}

\author{\fnms{Heng} \snm{Lian}
\ead[label=e3]{henglian@cityu.edu.hk}}
\address{Department of Mathematics\\
City University of Hong Kong\\
\printead{e3}}

\author{\fnms{Hua} \snm{Liang}
\ead[label=e4]{hliang@email.gwu.edu}}
\address{Department of Statistics\\
George Washington University\\
\printead{e4}}

\thankstext{t1}{Corresponding author}


\runauthor{B. Wang et al.}
\end{aug}

\begin{abstract}
We consider an additive partially linear framework for modelling massive heterogeneous data. The major goal is to extract multiple common features simultaneously across all sub-populations while exploring heterogeneity of each sub-population. We propose an aggregation type of estimators for the commonality parameters that possess the asymptotic optimal bounds and the asymptotic distributions as if there were no heterogeneity. This oracle result holds when the number of sub-populations does not grow too fast and the tuning parameters are selected carefully. A plug-in estimator for the heterogeneity parameter is further constructed, and shown to possess the asymptotic distribution as if the commonality information were available. Furthermore, we develop a heterogeneity test for the linear components and a homogeneity test for the non-linear components accordingly. The performance of the proposed methods is evaluated via simulation studies and an application to the Medicare Provider Utilization
and Payment data.
\end{abstract}


\tableofcontents
\end{frontmatter}

\section{Introduction}

Recent revolutions in technologies have produced many kinds of massive data, where the number of variables $p$ is fixed but the sample size $N$ is very large. In this paper we consider massive heterogeneous data. The analysis of non-massive heterogeneous data has been well studied in the literature. For example, non-massive heterogeneous data can be handled by fitting mixture models \citep{aitkin1985estimation} and by modeling variance functions \citep{davidian1987variance}. However, as far as we are aware, \cite{Zhao2016} is the only work that considers the analysis of massive heterogeneous data. They proposed a partially linear framework for modelling massive heterogeneous data, attempting to extract the common feature across all sub-populations while exploring heterogeneity of each sub-population. But the partially linear framework can only deal with only one common feature. In this paper, we propose an additive partially linear framework for modelling massive heterogeneous data, which can be applied to extract several common features across all sub-populations while exploring heterogeneity of each sub-population.

The additive partially linear models (APLMs) are a generalization of multiple linear regression models, and at the same time they are a special case of generalized additive nonparametric regression models \citep{hastie1990generalized}. As discussed in \cite{liu2011}, APLMs allow an easier interpretation of the effect of each variable and are preferable to completely nonparametric additive models, since they combine both parametric and nonparametric components when it is believed that the response variable depends on some variables in a linear way but is non-linearly related to the remaining independent variables. Estimation and inference for APLMs have been well studied in literature \citep[e.g., ][]{carroll2003semiparametric, opsomer1999root}. Recently, \cite{fang2015variance} proposed an approach for the analysis of heterogeneous data, fitting both the mean function and variance function using different additive partially linear models.

In this paper, we generalize the partially linear model (PLM) considered in \cite{Zhao2016} and propose an additive partially linear model (APLM) for modeling massive heterogeneous data. Let $\{(Y_i, \boldX_i, \boldZ_i)\}_{i=1}^N$ be the observations from a sample of $N$ subjects. We assume that there exist $s$ independent sub-populations, and the data from the $j$th sub-population follow the following additive partially linear model,
\begin{eqnarray} \label{eq:APLM}
Y^{(j)}=\boldX\trans \bbeta_0^{(j)} + \sum_{k=1}^K g_{0k}(Z_k) + \varepsilon, \mbox{\ \ for\ \ } j=1, \ldots, s,
\end{eqnarray}
where $\boldX=(X_1,\ldots,X_d)\trans$, $\boldZ=(Z_1, \ldots, Z_K)$, $\bbeta_0^{(j)}=(\beta_{01}^{(j)},\ldots, \beta_{0d}^{(j)})\trans$ is the vector of unknown parameters for $j$th sub-population, $g_{01},\ldots, g_{0K}$ are unknown smooth functions, and $\varepsilon$ has zero mean and variance $\sigma^2$.  Under model (\ref{eq:APLM}), $Y^{(j)}$ depends on $\boldX$ linearly but with coefficients varying across different sub-populations, whereas $Y^{(j)}$ depends on $\boldZ$ through additive non-linear functions that are common to all sub-populations. This model implies that the heterogeneity of the data is coming from the difference among $\bbeta_{0}^{(j)}$, $j=1, \ldots, s$.

Compared with \cite{Zhao2016}, the novelty of this paper is two-fold. First, we consider a more general and practical model. The partially linear model considered in \cite{Zhao2016} is a special case of (\ref{eq:APLM}) where $K=1$. Second, we use a different nonparametric tool (i.e., the regression splines tool) to fit the non-parametric functions than the reproducing kernel Hilbert Space (RKHS) tool that was used in \cite{Zhao2016}, making the theoretical development easier and computational implementation faster. The first fold of novelty is more significant than the second one, because it allows to consider more than one non-parametric components ($K>1$). However, the theoretical novelty of our approach is limited, because it would be straightforward generate the PLM in \cite{Zhao2016} to the APLM using the same RKHS tool. Instead, we propose to use the regression splines tool, because the resulting computational implementation is fast, especially for massive data. In order to understand this, let's count the number of knots when the RKHS tool and the regression splines are used, respectively. If the RKHS tool is used, the number of knots is $N$ for each non-parametric component, and therefore the total number of knots is equal to $KN$. If the regression splines is used, the number of knots is to be denoted as $J_N$, satisfying that $J_N\ll N$, and therefore the total number of knots is equal to $KJ_N$. The computational gain is small if the PLM is fitted, but the gain is significant if the APLM is fitted for massive data where $N$ is extremely large.

The rest of the paper is organized as follows. We propose the main methods in Section 2 and some hypothesis testing procedures in Section 3, deriving their asymptotic properties. We evaluate the performance of the proposed methods via simulation studies in Section 3 and a real data application in Section 4. We conclude the paper with a brief summary in Section 5 and relegate all the technical proofs to the Appendix.

\section{Methods}

\subsection{Notation and assumptions} \label{subsec:ind}

Recall that $\bbeta_0^{(j)}$ is the true sub-population specific parameter-vector for the $j$th sub-population, $j=1, \ldots, s$, and $g_0(\boldz)=g_{01}(z_1)+\cdots +g_{0K}(z_K)$ is the true additive common non-parametric function.  Without loss of generality, assume that $g_{0k}=g_{0k}(\cdot)$, $k=1, \ldots, K$, have a common support $[0,1]$. We propose to use polynomial splines \citep{carroll2003semiparametric} to approximate smooth function $g_{0k}$, $k=1, \ldots, K$. Let $\calS_N$ be the space of polynomial splines on $[0,1]$ of degree $\varrho \geq 1$, with a sequence of $J_N$ interior knots,
\begin{eqnarray*}
t_{-\varrho}=\cdots = t_{-1}=t_0=0<t_1<\cdots<t_{J_N} < 1=t_{J_N+1}=\cdots=t_{J_N + \varrho+1},
\end{eqnarray*}
where $J_N$ increases with the overall sample size $N$. Although we can choose different sequences of interior knots for different non-parametric functions in different sub-populations, for simplicity, as in \cite{liu2011}, here we consider the same sequence of equally spaced knots and let $h_N=1/(J_N +1)$ be the distance between neighboring knots.

Assume that $\boldX_{i}$ are i.i.d.~with $\boldX$ and $\boldZ_{i}$ are i.i.d.~with $\boldZ$. Define $\boldT=(\boldX,\boldZ)$. Let $m^{(j)}_0(\boldT)=\boldX\trans\bbeta^{(j)}_0 + g_0(\boldZ)$, $\Gamma(\boldz)=E(\boldX|\boldZ=\boldz)$, and $\wt{\boldX}=\boldX-\Gamma(\boldZ)$. And $\boldC^{\otimes2}$ denotes $\boldC\boldC\trans$ for any matrix or vector $\boldC$. Let $r$ be a positive integer and $\nu \in (0,1]$ such that $p=r+\nu > 2$. Let $\calH$ be the collection of functions $h$ on $[0,1]$ whose $r$th derivative exists and satisfies the Lipschitz condition of order $\nu$,
\begin{eqnarray*}
\left| h^{(r)}(z')-h^{(r)}(z) \right| \leq C |z'-z|^{\nu}, \forall \ 0\leq  z',z \leq 1,
\end{eqnarray*}
where and hereafter $C$ is a generic positive constant. In order to derive asymptotic results, we make the following mild assumptions.

\begin{enumerate}
\item[(A1).] Each component function $g_{0k} \in \calH, k=1,\ldots,K$ ;

\item[(A2).] The distribution of $\boldZ$ is absolutely continuous and its density $f$ is bounded away from zero and infinity on $[0,1]^K$;

\item[(A3).] There exists $c>0$ such that $c \| \bomega \|^2 \leq \bomega\trans E (\boldX^{\otimes 2}|\boldZ=\boldz) \bomega \leq C \| \bomega \|^2$, for any vector $\bomega \in {\mathcal R}^d$;

\item[(A4).] The number of interior knots $J_N$ satisfies: $N^{{1}/{(4p)}}\ll J_N \ll N^{{1}/{4}}$;

\item[(A5).] The projection function $\Gamma(\boldz)$ has the additive form $\Gamma(\boldz)=\Gamma_1(z_1)+\cdots +\Gamma_K(z_K)$, where $\Gamma_k \in \calH, E[\Gamma_k(z_k)]=0$ and $E[\Gamma_k(z_k)]^2 < \infty, k=1,\ldots,K$.
\end{enumerate}

In addition, to quantify the asymptotic consistencies of the non-parametric estimators, we consider both the empirical norms and the corresponding population norms. Let $\|\boldz\|$ be the Euclidean norm, $\|\boldz\|_\infty$ be the supremum norm, and $\|\boldz\|_1$ be the absolute-value norm of a vector $\boldz$, respectively. For a matrix $\boldC$, its $L_2$-norm is defined as $\|\boldC \|_2=\sup_{\|\boldu \| \neq 0} \|\boldC \boldu \|/\| \boldu\|$. Let $\| \varphi \|_{\infty} = \sup_{x \in [0,1]} | \varphi(x)|$ be the supremum norm of a function $\varphi$ on $[0,1]$. Following \cite{Stone1985} and \cite{Huang2003}, for any measurable function $\phi_1$ and $\phi_2$ on $[0,1]^K$, the empirical inner product and norm for the $j$th sub-sample and the whole sample, respectively, are defined as
\begin{eqnarray*}
\langle \phi_1, \phi_2 \rangle_{jn} = \frac{1}{n} \sum_{i \in \calG_j} \phi_1(\boldZ_i) \phi_2(\boldZ_i), \ \ \|\phi \|_{jn}^2=\frac{1}{n} \sum_{i \in \calG_j} \phi^2(\boldZ_i),\\
\langle \phi_1, \phi_2 \rangle_{N} = \frac{1}{N} \sum_{i=1}^N \phi_1(\boldZ_i) \phi_2(\boldZ_i), \ \ \|\phi \|_{N}^2=\frac{1}{N} \sum_{i=1}^N \phi^2(\boldZ_i).
\end{eqnarray*}
If $\phi_1$ and $\phi_2$ are $L^2$-integrable, the population inner product and norm are defined as
\begin{eqnarray*}
\langle \phi_1, \phi_2 \rangle = \int_{[0,1]^K} \phi_1(\boldz) \phi_2(\boldz) f(\boldz) d\boldz, \ \ \| \phi \|_2^2=\int_{[0,1]^K} \phi^2(\boldz)f(\boldz) d\boldz,
\end{eqnarray*}
where $f$ is the density of $\boldZ$. Similarly, for the $k$th component of $\boldZ$, $Z_k$ with density $f_k$, the empirical norm on the $j$th sub-sample, the empirical norm on the whole sample, and the population norm of any $L^2$-integrable univariate function $\varphi$ on $[0,1]$ are defined as
\begin{eqnarray*}
\| \varphi \|_{jnk}^2=\frac{1}{n} \sum_{i \in \calG_j} \varphi^2 (Z_{ik}), \ \ \| \varphi \|_{Nk}^2=\frac{1}{n} \sum_{i=1}^N \varphi^2 (Z_{ik}), \ \ \| \varphi \|_{2k}^2=\int_0^1 \varphi^2(z_k)f_k(z_k)d z_k.
\end{eqnarray*}

\subsection{Estimations for each sub-population} \label{subsec:ind}

First we consider the estimations for $\bbeta_0^{(j)}$ and $g_0=g_0(\cdot)$ based on the data from the $j$th sub-population only, $j=1, \ldots, s$. To this aim, let $G_j$ denotes the index set of all the observations from the sub-population $j$, and let $\calG_n^{(j)}=\{g^{(j)}(\cdot)\}$ be the collection of additive functions with the form that $g^{(j)}(\boldz)=g_1^{(j)}(z_1)+\cdots+g_K^{(j)}(z_K)$, where each component function $g_k^{(j)} \in \calS_N$ and $\sum_{i \in G_j} g_k^{(j)}(Z_{ik})=0$. Thus $\sum_{i \in G_j} g^{(j)}(\boldZ_i)=0$ for any $g^{(j)} \in \calG_n^{(j)}$.  For the $j$th sub-population, we consider the following estimators,
\begin{eqnarray}\label{eq:j-popu}
(\wh{\bbeta}^{(j)}, \wh{g}^{(j)}) =\argmin_{\stgbeta \in {\mathcal R}^d,\ g \in \calG_n^{(j)}}\left\{ L_n^{(j)}(\bbeta,g) = \frac{1}{2} \sum_{i \in G_j} \left[ Y_i - \boldX_i\trans \bbeta - g(\boldZ_i) \right]^2\right\}.
\end{eqnarray}

For the $k$th covariate $Z_k$, let $b_{m,k}(z_k)$ be the B-spline basis functions of degree $\varrho$ equipped with $J_N$ knots defined above. For any $g \in \calG_n^{(j)}$, we can write $g(\boldz)= \boldb(\boldz)\trans\bgamma$, where $\boldb(\boldz)=\{ b_{m,k}(z_k), m=-\varrho,\ldots,J_N, k=1,\ldots,K \}\trans$, which is a $K(J_N+\varrho+1)$-dim vector given $\boldz$, along with $K(J_N+\varrho+1)$-dim coefficient-vector $\bgamma=\{ \gamma_{m,k}, m=-\varrho,\ldots,J_N, k=1,\ldots,K \}\trans$. Therefore, (\ref{eq:j-popu}) is equivalent to
\begin{eqnarray}\label{eq:j-popu-equi}
\argmin_{\stgbeta \in {\mathcal R}^d,\ \stggamma \in {\mathcal R}^{K(J_N+\varrho+1)}}\left\{ l_n^{(j)}(\bbeta,\bgamma) = \frac{1}{2} \sum_{i \in G_j} \left[ Y_i - \boldX_i\trans \bbeta -  \boldb(\boldZ_i)\trans \bgamma\right]^2\right\},
\end{eqnarray}
if we consider the empirically centered estimator $\wh{g}^{(j)}(\boldz)=\sum_{k=1}^K\wh{g}^{(j)}_k(\boldz)$, where
\begin{eqnarray}\label{eq:center}
\wh{g}_k^{(j)} (z_k) = \sum_{m=-\varrho}^{J_N} \wh{\gamma}_{m,k} b_{m,k}(z_k) -\frac{1}{n} \sum_{i \in G_j} \sum_{m=-\varrho}^{J_N} \wh{\gamma}_{m,k} b_{m,k} (z_{ik}).
\end{eqnarray}

We derive some asymptotic results associated with the sub-population specific estimators, summarized in the following theorem.

\begin{Theorem}
\label{theorem1}
Under Assumptions (A1)-(A5), if the number of knots satisfies that $J_N \ll n^{1/2}$,
we have, for each sub-population, $j=1,\ldots,s$,
\begin{eqnarray*}
&&\|  \wh{g}^{(j)} - g_0 \|_2 = O_P\left( J_N^{{1}/{2}}n^{-{1}/{2}} + h_N^p \right)\\
&&\hspace{1cm} {\rm{\ \ and \ \ }} \| \wh{g}^{(j)} - g_0 \|_{jn} = O_P\left( J_N^{1/2}n^{-{1}/{2}} + h_N^p\right). \\
\end{eqnarray*}
If the number of knots further satisfies that $J_N \gg n^{1/(2p)}$
we have
\begin{eqnarray*}
\sqrt{n} \big(\wh{\bbeta}^{(j)}-\bbeta_0^{(j)}\big) \xrightarrow{d} \calN \big(\bfzero, \sigma^2 \boldD^{-1} \big),
\end{eqnarray*}
where $\boldD= E(\wt{\boldX}^{\otimes 2})$.
\end{Theorem}

\noindent {\bf Remark 1}: Assume that we consider $s=O(N^{1-\gamma})$ sub-samples, each sub-sample of $n=O(N^\gamma)$ observations, where $\gamma$ is some positive number between 0 and 1. In order to minimize the mean-square error of estimating $g_0$, $O_P(J_N^{1/2}n^{-1/2} + h_N^p)$, the best selection of $J_N$ is $O(N^{\frac{\gamma}{2p+1}})$, or equivalently, $O(n^{\frac{1}{2p+1}})$.  Under this selection, the mean-square error achieves the optimal rate, $O(N^{\frac{p\gamma} {2p+1}})$, or equivalently, $O(n^{\frac{p} {2p+1}})$.

\noindent {\bf Remark 2}: On the other hand, in order to ensure that $\wh{\bbeta}^{(j)}$ is $\sqrt{n}$-consistent for estimating $\bbeta_0^{(j)}$, we should adopt under-smoothing tuning with  $J_N \gg n^{1/(2p)}$ and carefully determine a balance between the number of sub-samples and the size of each sub-sample. For example, this can be achieved if we select $J_N$ as $O(N^q)$ with ${1}/{(4p)} < q <{1}/{4}$, and consider $s=O(N^{1-\gamma})$ sub-samples, each sub-sample of $n=O(N^{\gamma})$, with $2q < \gamma < 2pq$. The order of $J_N$ is consistent with the existing results in the literature. The recommended balance between $s$ and $n$ provides a guidance for the appropriate application of the divide-and-conquer strategy.

\subsection{Aggregation of commonality} \label{subsec:agg}

We consider the aggregated estimator, $\overline{g}(\boldz)=\frac{1}{s}\sum_{j=1}^s \wh{g}^{(j)}(\boldz)$, as the final estimator of $g_0(\boldz)$ based on the whole sample. Let $\calG_N$ be the collection of functions with the additive form $g(\boldz)=g_1(z_1)+\cdots+g_K(z_K)$, where $g_k \in \calS_N$ and $\sum_{j=1}^s\sum_{i \in G_j} g_k(Z_{ik})=0$. Thus, for any $g \in \calG_N$, $\sum_{j=1}^s\sum_{i \in G_j} g(\boldZ_i)=0$. In order to ensure that $\overline{g}\in \calG_N$, as in (\ref{eq:center}), we center the individual estimator $\wh{g}^{(j)}_k(z_k)$ via $\wh{g}_k^{(j)} (z_k) = \sum_{m=-\varrho}^{J_N} \wh{\gamma}_{m,k} b_{m,k}(z_k) -\frac{1}{N} \sum_{i=1}^N \sum_{m=-\varrho}^{J_N} \wh{\gamma}_{m,k} b_{m,k} (z_{ik})$. To abuse the notation, we still denote the centered estimator as $\wh{g}^{(j)}_k(z_k)$ and $\wh{g}^{(j)}(\boldz)=\sum_{k=1}^K \wh{g}^{(j)}_k(z_k)$. We derive the mean-square error of $\overline{g}$ in the following theorem.

\begin{Theorem} \label{theorem2}
Under Assumptions (A1)-(A5), if $J_N \ll n^{1/2}$, we have
\begin{eqnarray*}
\| \overline{g} - g_0 \|_2 = O_P \left( J_N^{1/2}N^{-1/2} + h_N^p \right), {\rm {\ \ and \ \ }} \| \overline{g} - g_0 \|_N = O_P \left( J_N^{1/2}N^{-1/2} + h_N^p  \right).
\end{eqnarray*}
\end{Theorem}

\noindent {\bf Remark 3}: In order to minimize the mean-square error of estimating $g_0$ using the aggregated estimator, if we select $J_N$ as $O(N^{\frac{1}{2p+1}})$, the mean-square error achieves the optimal rate $O(N^{\frac{p}{2p+1}})$.

\noindent {\bf Remark 4}: We compare the mean-square error of $\overline{g}$ with that of the following ``oracle estimator":
\begin{eqnarray*}
\wh{g}_{\rm oracle} = \argmin_{g \in \calG_N} \frac{1}{2} \sum_{j=1}^s \sum_{i \in G_j} \left[ Y_i - \boldX_i\trans \bbeta_0^{(j)} - g(\boldZ_i) \right]^2.
\end{eqnarray*}
assuming $\bbeta_0^{(j)}$, $j=1, \ldots, s$, are known. Following the proof of Theorem \ref{theorem1}, we can show that $\| \wh{g}_{\rm oracle} - g_0 \|_2 = O_P\left( J_N^{1/2} N^{-1/2}+h_N^p\right)$. Therefore, as long as $n \gg J_N^2$, the means-square errors of the aggregated estimator $\overline{g}$ and the oracle estimator $\wh{g}_{\rm oracle}$ are of the same order.

We conclude this subsection with some results for the massive homogeneous data where $\bbeta_0^{(j)} \equiv \bbeta_0, j=1, \ldots, s$. These results are of their own interest, when the divide-and-conquer strategy is applied to massive homogeneous data, where $\bbeta_0$ and $g_0$ are estimated using the aggregated estimators $\overline{\bbeta}=\frac{1}{s}\sum_{j=1}^s \wh{\bbeta}^{(j)}$ and $\overline{g}$, respectively. The result for $\overline{g}$ is the same as that in Theorem \ref{theorem2} and the result for $\overline{\bbeta}$ is stated in the following corollary.

\begin{Cor} \label{corollary1}
Consider homogeneous massive data where $\bbeta_0^{(j)} \equiv \bbeta_0, j=1,\ldots, s$. Under Assumptions (A1)-(A5), if $J_N \gg N^{1/(2p)}$ and $n \gg N^{1/2}$,
we have
$$ \sqrt{N} (\overline{\bbeta} - \bbeta_0 ) \xrightarrow{d} \calN \big( \bfzero, \sigma^2 \boldD^{-1} \big). $$
\end{Cor}

\subsection{Efficiency boosting for heterogeneous parameters} \label{subsec:boost}

The asymptotic variance matrix of $\wh{\bbeta}^{(j)}$ derived in Theorem \ref{theorem1} shows that there is some room to improve the estimation efficiency, because $\boldD^{-1}=E^{-1}(\wt{\boldX}^{\otimes 2})$ is bigger than the Cramer-Rao lower bound, $E^{-1}({\boldX}^{\otimes 2})$. Therefore, we re-substitute the aggregated estimator of $g$, $\overline{g}$, into \eqref{eq:j-popu} to improve the efficiency of estimating $\bbeta_0^{(j)}$. This leads to the following more efficient estimator,
\begin{eqnarray}
\breve{\bbeta}^{(j)} = \argmin_{\bbeta^{(j)} \in {\mathcal R}^d} \frac{1}{2} \sum_{i \in G_j} \left[Y_i - \boldX_i\trans \bbeta^{(j)} - \overline{g}(\boldZ_i) \right]^2.
\end{eqnarray}
for $j=1, \cdots, s$. We derive the asymptotic normality of $\breve{\bbeta}^{(j)}$ in the following theorem.

\begin{Theorem} \label{theorem3}
Under Assumptions (A1)-(A5), if $J_N$ satisfies the condition that $J_N\ll n^{1/2}$ given in the first part of Theorem \ref{theorem1} and the condition that $J_N\gg N^{1/(2p)}$ given in Corollary \ref{corollary1}, and it further satisfies that $J_N \ll s^{1/2}$, then we have
\begin{eqnarray*}
\sqrt{n} \big(\breve{\bbeta}^{(j)}-\bbeta_0^{(j)}\big) \xrightarrow{d} \calN \big(\bfzero, \sigma^2 \boldA^{-1} \big),
\end{eqnarray*}
where $\boldA= E({\boldX}^{\otimes 2})$.
\end{Theorem}

\noindent {\bf Remark 5}: As in Remarks 1-2, assume that we consider $s=O(N^{1-\gamma})$ sub-samples, each sub-sample of $n=O(N^\gamma)$ observations, where $\gamma$ is some positive number between 0 and 1. In order to satisfy all the conditions in Theorem \ref{theorem3}, we can consider $N^{2q} \ll n \ll N^{1-2q}$, with ${1}/{(2p)} < q < {1}/{4}$, and select $J_N=O(N^q)$. If $\boldX$ and $\boldZ$ are not independent, then $\boldA^{-1}< \boldD^{-1}$, implying that we can achieve such efficiency boosting through balancing between $n$ and $s$.

\subsection{Practical issues}

In this subsection, we consider several practical issues, including selection of the number of knots $J_N$, determination of linear components and non-linear components, and how to conduct statistical inference.

We first consider the selection of $J_N$. All the theoretical results need Assumption (A4): $N^{\frac{1}{4p}}\ll J_N \ll N^{\frac{1}{4}}$. Besides this, different theorem (or corollary) needs different an extra condition. Here is the list of those conditions:
\begin{eqnarray*}
&& {\rm (a)} \ \ J_N \ll n^{1/2}; \\
&& {\rm (b)} \ \ J_N \gg n^{1/(2p)}; \\
&& {\rm (c)} \ \ J_N \gg N^{1/(2p)}\mbox{\ and \ } n \gg N^{1/2}; \\
&& {\rm (d)} \ \ J_N \gg N^{1/(2p)}\mbox{\ and \ } J_N \ll s^{1/2}.
\end{eqnarray*}
In Theorem \ref{theorem1}, under Condition (a), we derive the bound for the mean-square error of each sub-population specific estimator $\wh{g}^{(j)}$, $j=1, \cdots s$. In Theorem \ref{theorem1}, under Conditions (a) and (b), we derive the asymptotic normality for each sub-population specific estimator $\wh{\bbeta}^{(j)}$, $j=1, \cdots s$. In Theorem \ref{theorem2}, under Condition (a), we derive the bound for the mean-square error of the aggregated estimator $\overline{g}$. In Corollary \ref{corollary1}, under Condition (c) and for the massive homogeneous data, we derive the asymptotic normality for the aggregated estimator $\overline{\bbeta}$. In Theorem \ref{theorem3}, under Condition (d), we derive the asymptotic normality for each sub-population specific efficiency-boosted estimator $\breve{\bbeta}^{(j)}$. These conditions can be satisfied by carefully selecting the balance between $n$ and $s$, with some guidance provided in Remarks 1-5.

However, it is hard to use these theoretical requirements on the order of $J_N$ to guide the selection of $J_N$ in practice. If computational power allows, we can utilize cross-validation to select $J_N$ adaptively. If cross-validation is used, as discussed in \cite{xu2016optimal}, we should consider distributed cross-validation aiming for global optimality (selecting a single $J_N$ based on the entire dataset) instead of subsample cross-validation aiming for local optimality (selecting an $J_N$ for each subsample separately). As we consider massive data here, we don't recommend any data-driven tuning procedure, which requires heavy computational cost. Instead, as in most studies using regression splines, we recommend the fixed choice of the number of internal knots $J_N$ (e.g., some small number between 2 and 10). Although pre-specifying a fixed $J_N$ which might be sub-optimal, it is much more convenient and computationally efficient than any data-driven procedure. This is another advantage of the regression splines compared with smoothing splines: the number of knots in the regression splines can be determined directly in terms of model construction without looking at the data, while in the smoothing splines the tuning parameter $\lambda$, which controls the balance between goodness-of-fit and function smoothness, has to be determined adaptively. In order to determine tuning parameter $\lambda$ in the smoothing splines, we have to use data-driven procedure as in \cite{xu2016optimal}.

Another practical issue is the determination of linear components and non-linear components. In practice, we rely on graphical tools such as box-plot and scatterplot to have a rough idea on which components might be linear or non-linear. For the setting in this paper, we consider parametric function for each heterogeneous component (the sample size is smaller for each subsample, so a simpler model is considered) and non-parametric function for each homogeneous component (the sample size is larger for the entire sample, so a more flexible model is considered). Motivated by \cite{lu2016} and \cite{Zhao2016}, some formal hypothesis testing methods for heterogeneity and homogeneity are proposed in the next section.

The third practical issue is conducting statistical inferences about the regression parameters. In the above theorems, we derive the explicit formulas for their corresponding covariance matrices by examining the asymptotic normality. If the covariance matries were to involve the density of error term, we wouldn't apply the plug-in procedure to estimate them as discussed in Section 4 of \cite{volgushev2017distributed}. Fortunately, all the asymptotic covariance matrices we developed in the above theorems can be estimated using the plug-in procedure and conducting statistical inferences is straightforward. In addition, as discussed in Remark 2, in order to achieve the estimation efficiency of regression parameters, the non-parametric functions should be under-smoothing. This is another reason that we don't recommend data-driven method for the determination of $J_N$. If $J_N$ is selected adaptively, the estimation of regression parameters may be sub-efficient and it is also subject to the problem of inference-after-selection.

\section{Hypothesis testing}

\subsection{Testing heterogeneity} \label{sec:test_hetero}

As in \cite{Zhao2016}, we develop statistical tests for the heterogeneity of the linear components across sub-populations. First, we consider a general class of pairwise test for heterogeneous parameters. Then, we develop a more general heterogeneity test involving many (up to as many as $s$) sub-populations.

First, consider the following general class of pairwise test for heterogeneous parameters:
\begin{eqnarray} \label{test:H0}
H_0: \ \boldQ \big( \bbeta_0^{(j_1)} - \bbeta_0^{(j_2)}\big) = \bfzero \quad {\rm vs.} \quad H_a: \ \boldQ \big( \bbeta_0^{(j_1)} - \bbeta_0^{(j_2)}\big) \neq \bfzero,
\end{eqnarray}
where $j_1 \neq j_2 \in \{ 1,\ldots, s\}$, and $\boldQ=(\boldq_1\trans,\ldots,\boldq_{d_1}\trans)\trans$ is a $d_1 \times d$ matrix with $d_1\leq d$. This class of tests includes testing if either the whole vector or specific entries of $\bbeta_0^{(j_1)}$ are equal to those of $\bbeta_0^{(j_2)}$. It is straightforward to consider $\boldQ \big( \wh\bbeta^{(j_1)} - \wh\bbeta^{(j_2)}\big)$ or $\boldQ \big( \breve\bbeta^{(j_1)} - \breve\bbeta^{(j_2)}\big)$ as test statistic, which is based on the estimators from Subsection 2.2 or the estimators from Subsection 2.4, respectively. We summarize the asymptotic properties of these two test statistics in the following theorem, based on which we can conduct the Wald tests.

\begin{Theorem} \label{theorem4}
If the conditions in Theorem \ref{theorem1} hold, under the null hypothesis \eqref{test:H0},
$$ \sqrt{n} \boldQ \big( \wh\bbeta^{(j_1)} - \wh\bbeta^{(j_2)}\big) \xrightarrow{d} \calN \big( \bfzero, 2\sigma^2 \boldQ \boldD^{-1} \boldQ\trans \big).$$
Furthermore, if the conditions in Theorem \ref{theorem3} hold, under the null hypothesis \eqref{test:H0},
$$ \sqrt{n} \boldQ \big( \breve\bbeta^{(j_1)} - \breve\bbeta^{(j_2)}\big) \xrightarrow{d} \calN \big( \bfzero, 2\sigma^2 \boldQ \boldA^{-1} \boldQ\trans \big).$$
\end{Theorem}

Based on Theorem \ref{theorem4}, we can construct the Wald tests as what follows,
\begin{eqnarray*}
\Psi_1 &=& I\left\{ \boldQ \big( \wh\bbeta^{(j_1)} - \wh\bbeta^{(j_2)}\big) \notin \sqrt{\frac{2}{n}} \sigma (\boldQ \boldD^{-1} \boldQ\trans)^{\frac{1}{2}}Z_{1-\alpha/2} \right\}, \\
\Psi_2 &=& I\left\{ \boldQ \big( \breve\bbeta^{(j_1)} - \breve\bbeta^{(j_2)}\big) \notin \sqrt{\frac{2}{n}} \sigma (\boldQ \boldA^{-1} \boldQ\trans)^{\frac{1}{2}}Z_{1-\alpha/2} \right\},
\end{eqnarray*}
where $\alpha$ is a given significant level and $Z_{1-\alpha/2}$ is the $(1-\alpha/2)$ quantile of a standard normal distribution. It is clear that $\Psi_2$ is more powerful than $\Psi_1$ because of a smaller variance of $\breve\bbeta^{(j)}$ compared with $\wh\bbeta^{(j)}$. Coverage probabilities, length of confidence intervals, nominal levels and powers are empirically evaluated in simulation studies, which confirm our theoretical results.

Then, we consider a more general heterogeneity test involving many sub-populations indexed by $j, j \in \calS \subseteq \{1,\ldots,s\}$. Note that $|\calS|$ is allowed to be as large as $s$, i.e., all sub-populations. The null and alternative hypotheses are formulated as
\begin{eqnarray} \label{test:H0_multiple}
H_0: \ \boldQ \bbeta_0^{(j)} = \boldQ \check\bbeta_0^{(j)} \ \ \forall j \in \calS \quad {\rm vs.} \quad H_a: \boldQ \ \bbeta_0^{(j)} \neq \boldQ \check\bbeta_0^{(j)} \ \ \exists j \in \calS,
\end{eqnarray}
where $\check\bbeta_0^{(j)}$'s are some predetermined values.

To test \eqref{test:H0_multiple}, it is natural to define the following test statistic:
\begin{eqnarray*}
T_{\calS} = \max_{j \in \calS} \sqrt{n} \| \boldQ (\breve\bbeta^{(j)} - \check\bbeta_0^{(j)} ) \|_\infty.
\end{eqnarray*}
Its distribution can be approximated by bootstrapping the quantity
\begin{eqnarray*}
W_\calS =  \max_{j \in \calS} \frac{1}{\sqrt{n}} \left\| \sum_{i\in G_j} \boldQ (\wh\boldA_n^{(j)})^{-1} \boldX_i e_i \right\|_\infty,
\end{eqnarray*}
where $\wh\boldA_n^{(j)}=\frac{1}{n}\sum_{i \in G_j} \boldX_i^{\otimes 2}$ and $e_i \sim \calN (0,\sigma^2)$'s are i.i.d. The following theorem summarizes the consistency of the proposed multiplier bootstrap method.

\begin{Theorem} \label{theorem5}
Suppose the conditions in Theorem \ref{theorem3} are satisfied. Under the null hypothesis \eqref{test:H0_multiple}, for any $\calS \subseteq \{1,\ldots,s\}, |\calS|=d$, if $s \gg J_N^2 \log(pd)$, $(\log(pdn))^7/n \leq c_1 n^{-c_2}$ for some constants $c_1,c_2 >0$, and $p^2 \log(pd)/\sqrt{n}=o(1)$, then we have
$$ \sup_{\alpha \in (0,1)} \left| P\big( T_\calS > c_\calS(\alpha) \big) - \alpha \right| = o(1), $$
where $c_\calS=\inf \{w \in \mathcal{R} : P(W_\calS \leq w | \boldX ) \geq 1-\alpha\}.$
\end{Theorem}

\noindent {\bf Remark 6}: Actually, the above hypotheses can be extended to test the heterogeneity of all sub-populations without specifying $\check\bbeta_0^{(j)}$'s. With a similar argument as that in \cite{Zhao2016}, we only need to test if the differences of heterogeneity parameters between any two consecutive sub-populations equal to zero. The modified test statistic is
\begin{eqnarray*}
T'_{\calS} = \max_{1 \leq j \leq s-1} \sqrt{n} \| \boldQ ( \breve\bbeta^{(j)} - \breve\bbeta^{(j+1)}) \|_\infty,
\end{eqnarray*}
and corresponding bootstrap quantity is given by
\begin{eqnarray*}
W'_\calS =  \max_{1 \leq j \leq s-1} \frac{1}{\sqrt{n}} \left\| \sum_{i\in G_j} \boldQ (\wh\boldA_n^{(j)})^{-1} \boldX_i e_i - \sum_{i\in G_{j+1}} \boldQ (\wh\boldA_n^{(j+1)})^{-1} \boldX_i e_i\right\|_\infty.
\end{eqnarray*}
Similarly, we define $c'_\calS=\inf \{w \in \mathcal{R} : P(W'_\calS \leq w | \boldX ) \geq 1-\alpha\}.$

\subsection{Testing homogeneity} \label{sec:test_homo}

Now we consider the test of whether the non-linear components, $g_{0k}(Z_k), k=1,2,\ldots, K$ are homogeneous across all $s$ sub-populations, which is the necessity of doing aggregation of commonality. With a little abuse of notation, for the $j$th sub-population, we denote the true unknown smooth functions as $g_{0k}^{(j)}(Z_k),k=1,\ldots, K$. We apply the likelihood ratio principle to the following homogeneity test for the $k$th smooth function,
\begin{eqnarray} \label{test:H0_homo}
H_0: \ g_{0k}^{(1)}(Z_k) = \cdots = g_{0k}^{(s)}(Z_k) \quad {\rm vs.} \quad H_a: g_{0k}^{(j)}\rm{'s} \mbox{\ are\ not\ all \ the \ same}.
\end{eqnarray}

Let $g_{-k} = \sum_{k'\neq k} g_{k'}$ and $\boldZ_{-k}=(Z_1,\ldots,Z_{k-1},Z_{k+1},\ldots,Z_K)$. The subscript $-k$ means removing the $k$th component therein. Define
$$L_{nk}^{(j)}(\bbeta,g_k,g_{-k}) = \frac{1}{2n} \sum_{i \in G_j} \left[ Y_i - \boldX_i\trans \bbeta - g_k(Z_k) - g_{-k}(\boldZ_{i,-k}) \right]^2.$$
We can construct a likelihood ratio test statistic as
\begin{eqnarray}
{\rm LRT}_{nk}^s = \sum_{j=1}^{s-1} \left\{ L_{nk}^{(j)}(\wh\bbeta^{(j)},\wh{g}_k^{(j)},\wh{g}_{-k}^{(j)}) - L_{nk}^{(j)}(\wh\bbeta^{(j)},\wh{g}_k^{(j+1)},\wh{g}_{-k}^{(j)}) \right\}.
\end{eqnarray}

Before providing the limiting distribution for the above test, extra notations are required.
We say that a statistic $T_n$ is nearly $\chi^2_{b_n}$, denoted as $T_n \stackrel{a}{\sim} \chi^2_{b_n}$, if $(2b_n)^{-1/2}(T_n-b_n)$ weakly converges to $\mathcal{N}(0,1)$ for some sequence $b_n \to \infty$.

\begin{Theorem} \label{theorem6}
Suppose the conditions in Theorem \ref{theorem1} are satisfied. Under the null hypothesis \eqref{test:H0_homo}, if $J_N^{1/2} = o(s)$, we have
$$ -\frac{1}{3\sigma^2}n \cdot {\rm LRT}_{nk}^s \stackrel{a}{\sim} \chi^2_{u_N},$$
where $u_N=\frac{2}{3}(s-1) J_N$.
\end{Theorem}

\noindent {\bf Remark 7}: $J_N^{1/2} = o(s)$ is a weak requirement, implying the number of sub-populations cannot be too small to borrow sufficient information across all of them. This phenomenon is also observed in \cite{lu2016}, in which it is called the ``blessing of aggregation". According to Remarks 1 and 2, $J_N =O(N^q)$, where $1/(4p) < q<1/4$, implying that $J_N^{1/2} = o(s)$ means $\gamma < 1-q/2$ with the minimum value of $7/8$ of the right hand side.

\noindent {\bf Remark 8}: The above theorem considers testing one smooth function. If we want to test the summation of all smooth functions, the result can be proved similarly, which is summarized in the following theorem by removing the extra condition $J_N^{1/2} = o(s)$ due to no usage of both $\wh{g}_k^{(j+1)}$ and $\wh{g}_{-k}^{(j)}$ in the $L_{nk}^{(j)}$.

\begin{Theorem} \label{theorem7}
Suppose the conditions in Theorem \ref{theorem1} are satisfied. Under the null hypothesis $H_0: \ g_{0}^{(1)}(\boldZ) = \cdots = g_{0}^{(s)}(\boldZ)$,
we have
$$ -\frac{1}{3\sigma^2 }n \cdot {\rm LRT}_{nk}^s \stackrel{a}{\sim} \chi^2_{u_N},$$
where $u_N=\frac{2}{3}(s-1) KJ_N$.
\end{Theorem}

\section{Simulation Studies}

We conduct simulation studies to examine the impact of the balance between sub-population sizes $n$ and the number of sub-population $s$ on the performance of the proposed estimators, $\overline{g}$ and $\breve{\bbeta}^{(j)}$. We consider the following additive partially linear model with two nonparametric components ($K=2$) as the data generating model:
\begin{eqnarray*}
Y^{(j)} &=& X\beta_0^{(j)} + g_1(Z_1) + g_2(Z_2) + \varepsilon, \\
g_1(Z_1) &=& 5 \sin \{2\pi(Z_1+1)\}, \\
g_2(Z_2) &=& 100 \left( e^{-1.625(Z_2+1)} - 4 e^{-3.25(Z_2+1)} + 3 e^{-4.825(Z_2+1)} \right) - C_0,
\end{eqnarray*}where $\varepsilon$ is generated from normal distribution $N(0,1)$,  $Z_1$, $Z_2$ and $W$ are generated independently from uniform distribution $U(-1,1)$, $X=\frac{1}{2}(W+Z_1)$, and  $C_0$ is taken as $100(1-e^{-3.25})/3.25 - 400(1-e^{-6.5})/6.5+300(1-e^{-9.75})/9.75$ to make sure that $E\{g_1(Z_1)\}=E\{g_2(Z_2)\}=0$. We can show that $\wt{X}=W/2$, $\boldD=E(\wt{X}^2) = 1/12$, and $\boldA=E(X^2)=1/6$. In order to generate heterogenous data, we let $\beta_0^{(j)}=j$, for the $j$th sub-population, $j=1,\ldots,s$, with $d=1$.

In order to $g_1$ and $g_2$ using polynomial splines, we consider cubic splines ($\varrho=3$) and equal-spaced knots. We estimate the unknown error variance $\sigma^2$ using $\overline{\sigma}^2 = \sum_{j=1}^s ( \wh{\sigma}^{(j)})^2 / s$, where
\begin{eqnarray*}
(\wh{\sigma}^{(j)})^2 = \frac{1}{n-d-K(J_N +\varrho)} \sum_{i \in G_j} \left[ Y_i -X_i \wh\beta^{(j)} - \wh{g}^{(j)}(\boldZ_{i}) \right]^2.
\end{eqnarray*}
We set the massive sample size $N$ as $2^{11}, 2^{12}, 2^{13}$, or $2^{14}$. We set the number of sub-samples $s$ as $N^{1-\gamma}$, where $\gamma=\max(0.4,2q),\ldots, 0.9, 1$. We set the minimal value of $\gamma$ as $\max(0.4,2q)$ to ensure that $J_N^2=O(N^{2q}) \ll n=O(N^{\gamma})$. For each setting, we run 200 repetitions.

\begin{figure}[htb]
\protect\caption{Root mean-square-errors of the aggregated estimator, $\overline{g}$, under different settings of the number of knots, the number of sub-samples, and the sample size.}
\centering \includegraphics[scale=0.5]{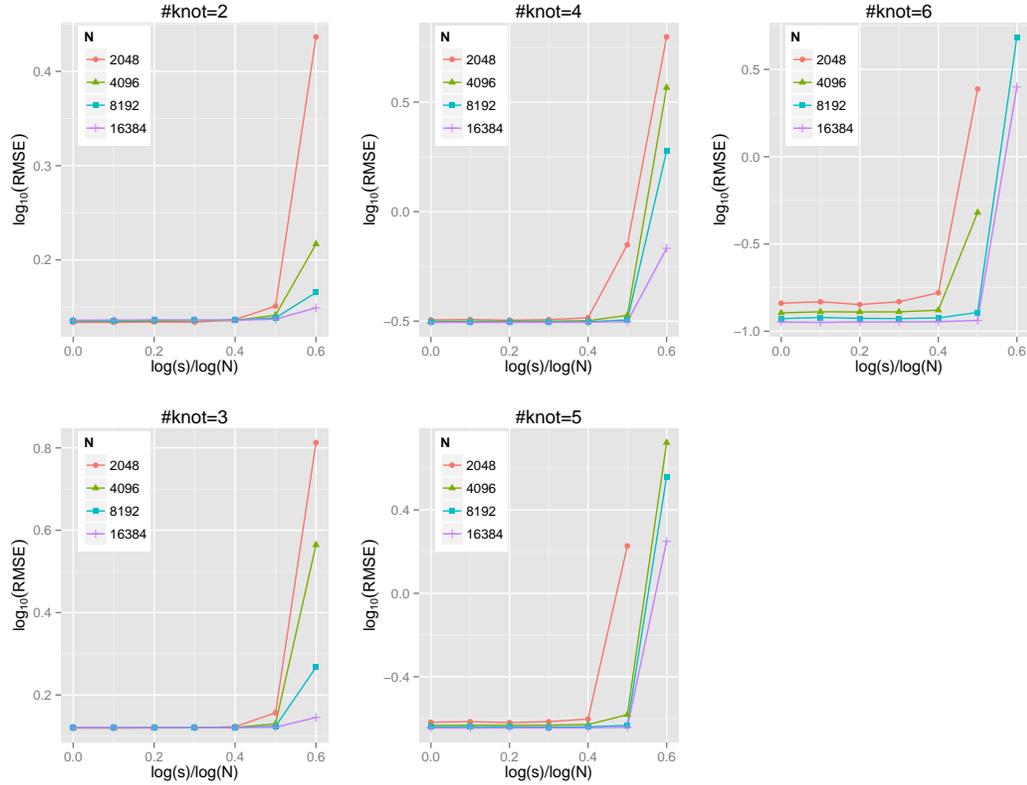}\label{fig:figure1}
\end{figure}

First, we evaluate the performance of the aggregated estimator, $\overline{g}$, as an estimator for $g$. We compute the root mean-square-error (RMSE) of $\overline{g}$, under different choices of $J_N$ and $s$, and different settings of $N$. The results are summarized in Figure \ref{fig:figure1}. The condition that $J_N^2 \ll n$, which is needed in all the theorems, implies that the larger number of knots we take and the shorter range of $s$ we should consider. In Figure \ref{fig:figure1}, for each selection of the number of knots, we see that the performance of $\overline{g}$ is good and stable during a wide range of $s$. We also see that the RMSE of $\overline{g}$ deteriorates quickly when $\log(s)/\log(N)$ is approaching $1-2q, q \approx \log_N(J_N)$. For example, using 5 knots, $N=2^{11}, q=\log_N(5) \approx 0.21$ and then $1-2q \approx 0.42$; therefore, from the second figure in the bottom row of Figure \ref{fig:figure1}, we see that corresponding RMSE increases a lot when the ratio approaches $0.5$. In summary, from \ref{fig:figure1}, we see there is a clear boundary of $\log(s)/\log(N)$: with this boundary, the performance of $\overline{g}$ is very good, while beyond this boundary, the performance is very bad.  These findings confirm the theoretical results presented in Theorem \ref{theorem2}.

\begin{figure}[!htb]
\protect\caption{Coverage probabilities and interval lengths of 95\% confidence intervals, CI$_1$ and CI$_2$, under different settings of the number of knots and the number of sub-samples, with $N=2^{11}$.}
\centering\includegraphics[scale=0.8]{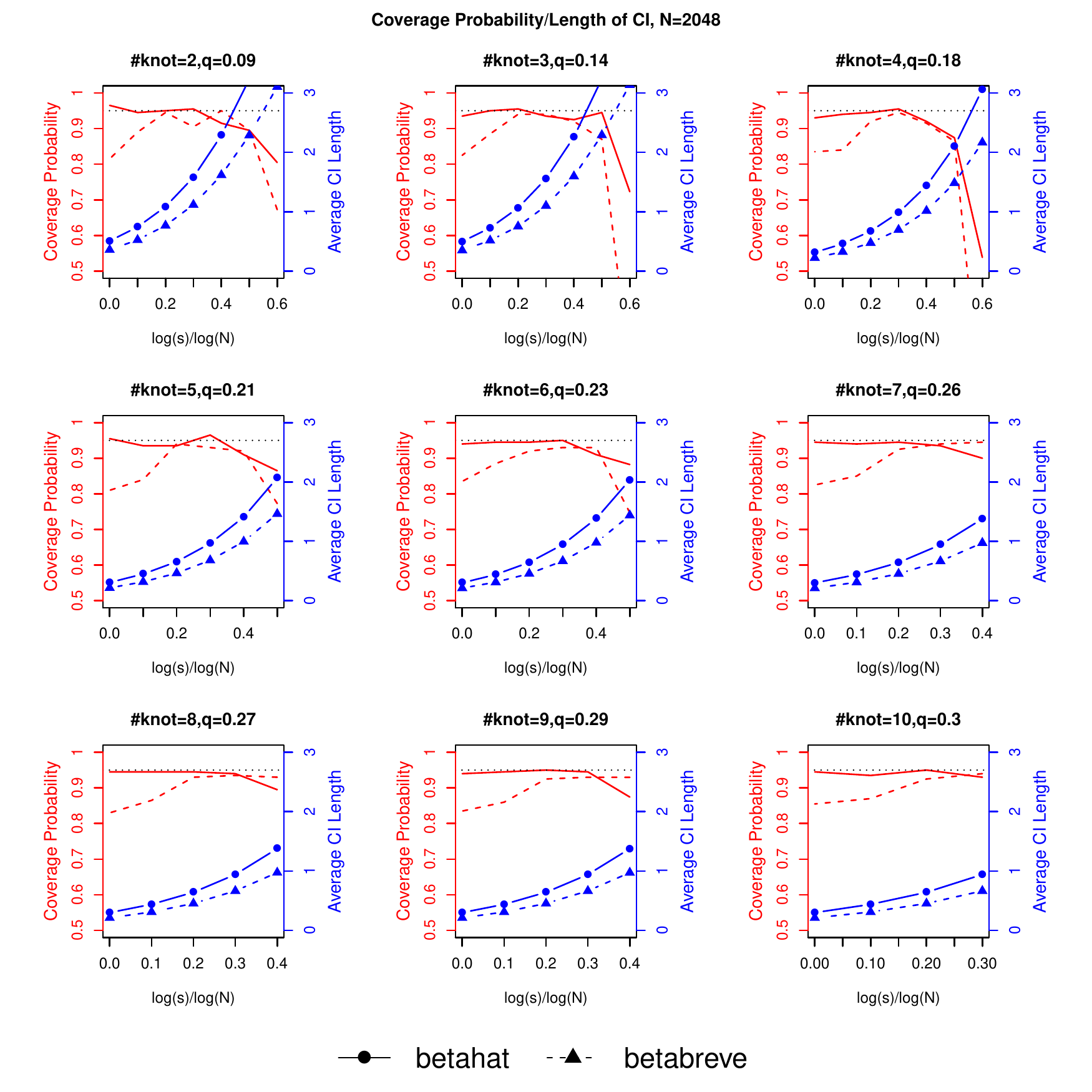}\label{fig:figure2}
\end{figure}

\begin{figure}[!htb]
\protect\caption{Coverage probabilities and interval lengths of 95\% confidence intervals, CI$_1$ and CI$_2$, under different settings of the number of knots and the number of sub-samples, with $N=2^{14}$.}
\centering \includegraphics[scale=0.8]{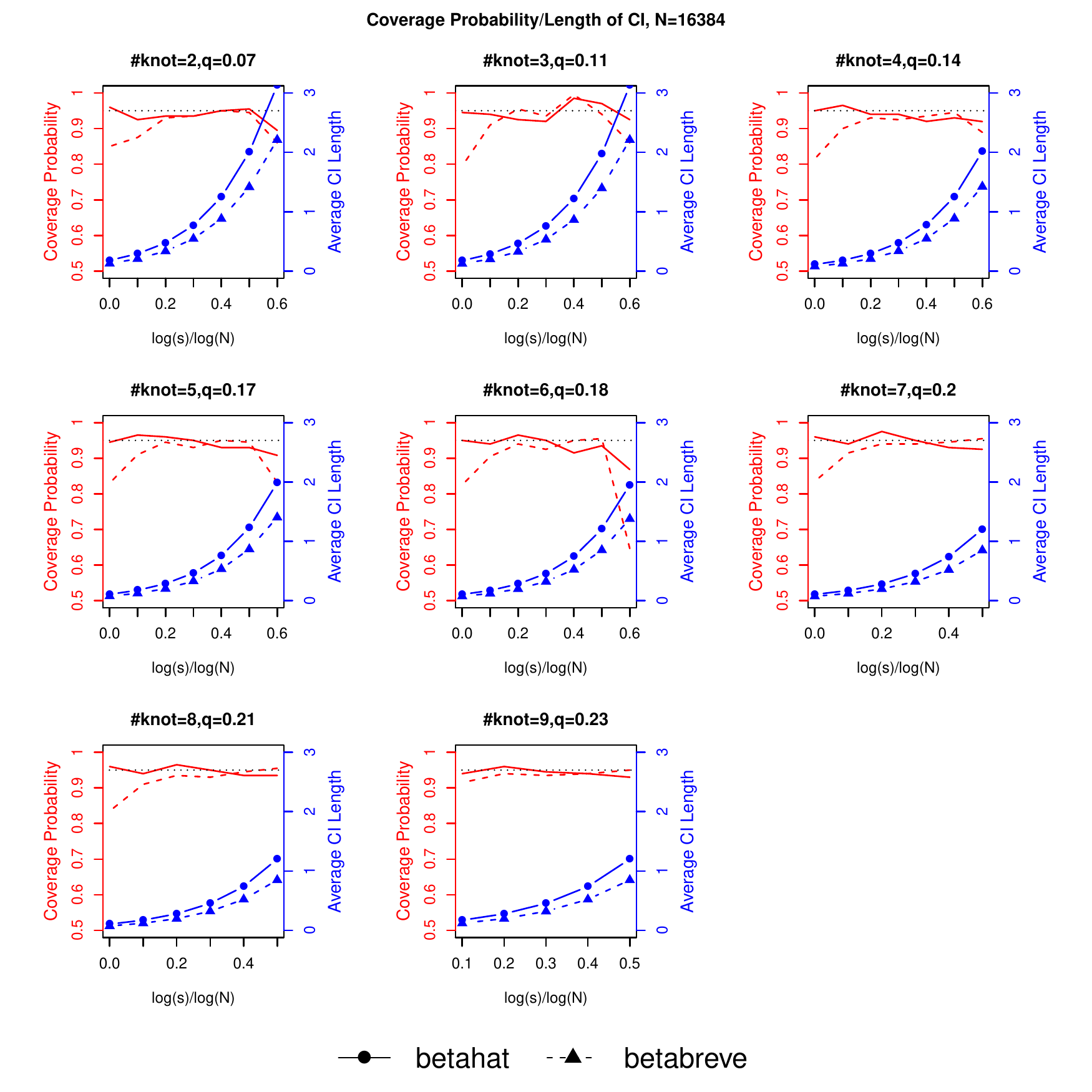}\label{fig:figure3}
\end{figure}

Second, we evaluate the performance of the proposed estimators, $\wh{\bbeta}^{(j)}$ and $\breve{\bbeta}^{(j)}$, for estimating $\beta_0^{(j)}$. We consider $95\%$ confidence intervals based on $\wh{\bbeta}^{(j)}$ and $\breve{\bbeta}^{(j)}$ respectively as follows:
\begin{eqnarray*}
\textrm{CI}_1  = \left[ \wh{\bbeta}^{(j)} \pm \frac{1.96 \overline{\sigma}}{\sqrt{n}} \boldD^{-1/2} \right] {\rm{\ \  and \ \ }}\textrm{CI}_2  = \left[ \breve{\bbeta}^{(j)} \pm \frac{1.96 \overline{\sigma}}{\sqrt{n}} \boldA^{-1/2} \right].
\end{eqnarray*}
For simplicity, we summarize results for the first sub-population in Figures 2-4, where both the coverage probabilities and the interval lengths are displayed, with the results of  $\wh{\beta}^{(1)}$ in solid line with circle and those of $\breve{\beta}^{(1)}$ in dashed line with triangle.

\begin{figure}[!htb]
\protect\caption{Coverage probabilities of 95\% CI$_2$ confidence intervals under different settings of the number of knots, the number of sub-samples, and the sample size.}
\centering \includegraphics[scale=0.5]{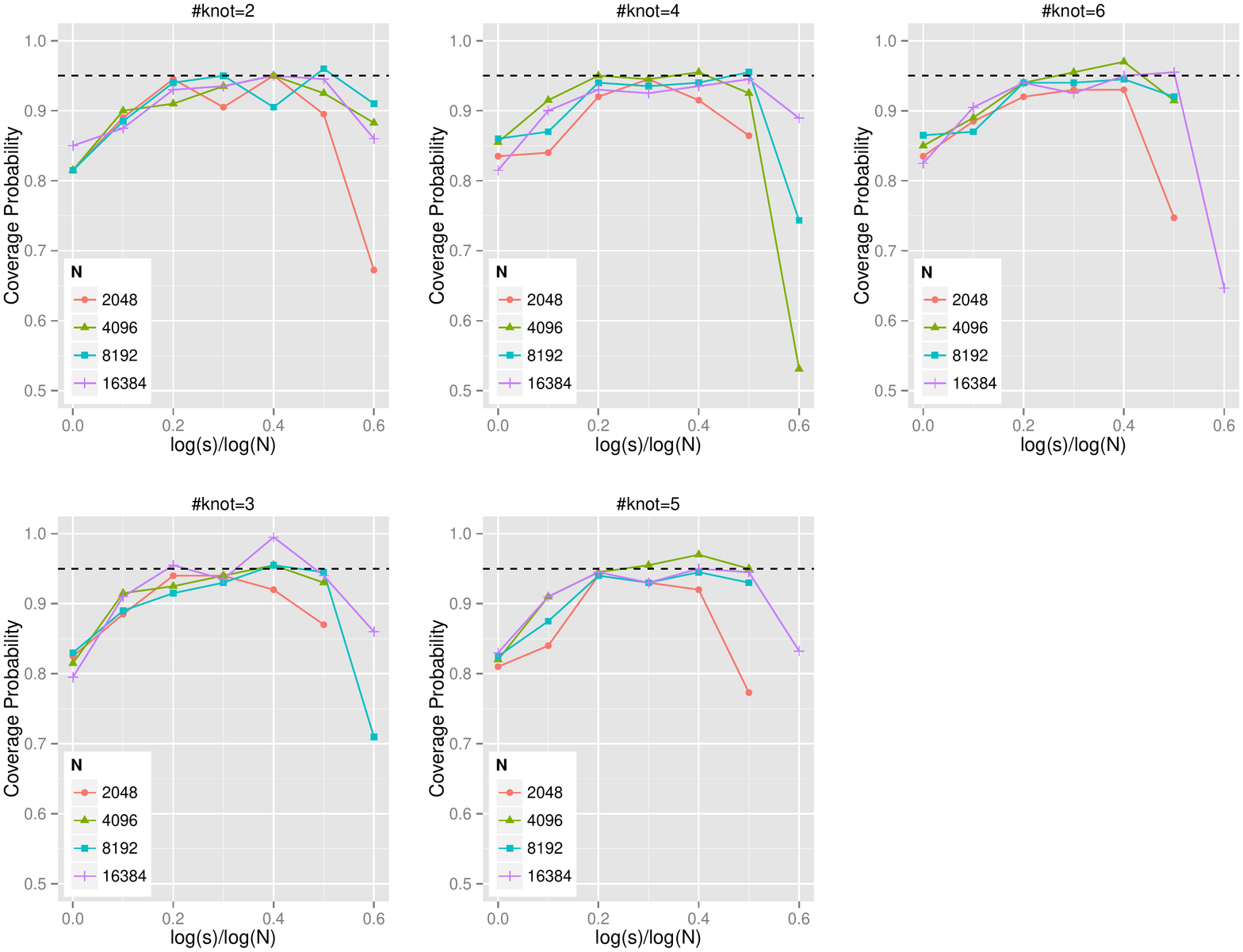}\label{fig:figure4}
\end{figure}

From Figure 2 where $N=2^{11}$ and Figure 3 where $N=2^{14}$, we see that within a proper range of $s$, CI$_1$ and CI$_2$ have similar coverage probabilities. We also see that on average, the interval length of CI$_2$ is shorter than that CI$_1$. This finding confirm that the asymptotic variance derived in Theorem \ref{theorem3} is smaller than that in Theorem \ref{theorem1}. However, the coverage probability of CI$_2$ is valid for a shorter range of $\log(s)/\log(N)$, in contrast with that of CI$_1$. This is finding is consistent with that there are more conditions in Theorem \ref{theorem3} than in Theorem \ref{theorem1}.

To visualize the performance of CI$_2$ more clearly, in Figure \ref{fig:figure4} we display the coverage probability of CI$_2$ in more detail for different settings of $s$ and $N$, given different numbers of knots. From Figure \ref{fig:figure4}, we can see that, given the number of knots, a larger $N$ implies a wider valid range for $s$ to achieve a good coverage; given $N$, a larger number of knots implies a smaller transition point for $s$.

Third, we evaluate the heterogeneity tests using the following Wald test statistics constructed based on Theorem \ref{theorem4}:
\begin{eqnarray*}
\Psi_1 &=& I\left\{ \boldQ \big( \wh\bbeta^{(j_1)} - \wh\bbeta^{(j_2)}\big) \notin \sqrt{\frac{2}{n}} \overline{\sigma} (\boldQ \wh\boldD^{-1} \boldQ\trans)^{1/2}C_{\alpha/2} \right\}, \\
\Psi_2 &=& I\left\{ \boldQ \big( \breve\bbeta^{(j_1)} - \breve\bbeta^{(j_2)}\big) \notin \sqrt{\frac{2}{n}} \overline{\sigma} (\boldQ \wh\boldA^{-1} \boldQ\trans)^{1/2}C_{\alpha/2} \right\},
\end{eqnarray*}
where $C_{\alpha/2}$ is the upper $\alpha/2$ quantile of a standard normal distribution, and $\wh\boldD$ and $\wh\boldA$ are the sample estimators of $\boldD$ and $\boldA$, respectively. The results are summarized in Figure \ref{fig:figure5}, where $\Psi_1$ and $\Psi_2$ are compared in terms of Type-I error and power, under different settings of $s$ and $N$. From Panel (a) of Figure \ref{fig:figure5}, we see that both $\Psi_1$ and $\Psi_2$ have appropriate type-I error within a wide range of $s$, but they have inflated type-I error after $s$ passes a transition point. Panels (b)-(d) compare the testing powers under three different alternative hypotheses: $H_a: \beta_0^{(j_1)} - \beta_0^{(j_2)} = \Delta$, where $\Delta=0.5, 1$ and $1.5$, respectively. We see that the power increases as $N$ increase and $\Delta$ increases. We also see the power of $\Psi_2$ is larger than that of $\Psi_1$ across different settings.  These findings confirm the asymptotic results stated in Theorem \ref{theorem4}.

\begin{figure}[!htb]
\protect\caption{Type-I error and power of tests $\Psi_1$ and $\Psi_2$ under different settings of the number of sub-samples and the sample size, using 4 knots.}
\centering \includegraphics[scale=0.6]{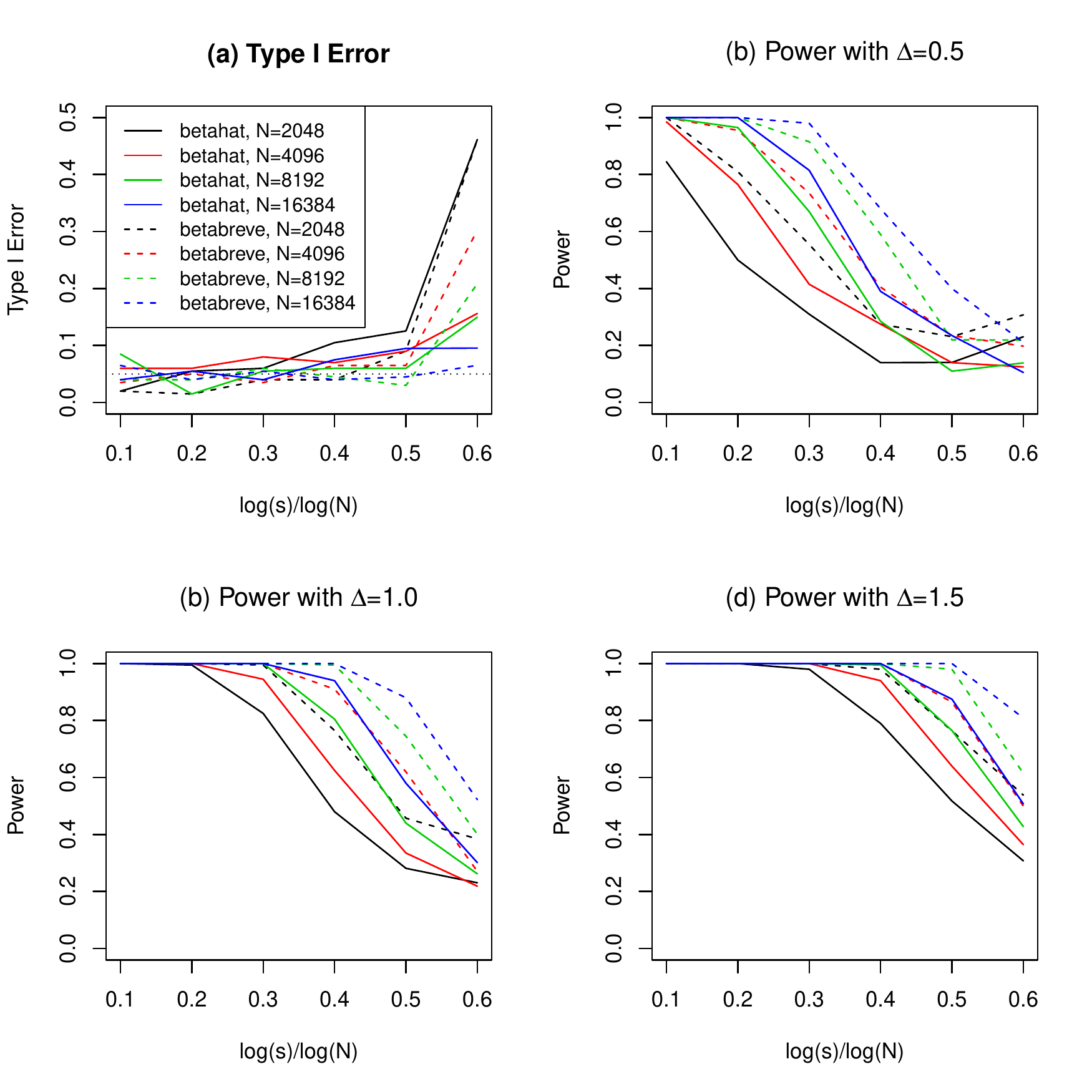}\label{fig:figure5}
\end{figure}

Fourth, we evaluate the more general heterogeneity tests involving all sub-populations using the test statistics constructed based on Theorem \ref{theorem5} and Remark 6. The null and alternative hypotheses are formulated as follows:
\begin{eqnarray*}
H_0: \ \beta_0^{(1)} = \cdots =\beta_0^{(s)} \quad {\rm vs} \quad H_a: \beta_0^{(1)} =  \beta_0^{(j)} + \Delta  \ \ j =2,\ldots,s,
\end{eqnarray*}
where $\Delta=0, 0.4, 0.6$ and $1$. Note that here $\boldQ=1$. In the simulations, we set $\beta_0^{(j)}=1$ for the null hypothesis, and use 500 bootstrapping samples to obtain the upper $\alpha=0.05$ quantile for $W'_\calS$. The results are summarized in Figure \ref{fig:figure6}, where the $y$-axis shows the probability $P(T'_{\calS} > c'_{\cal}S)$, under different settings of $s$ and $N$. From Panel $\Delta=0$ of Figure \ref{fig:figure6}, we see that the proposed test statistic $T'_{\calS}$ has appropriate type-I error around the nominal level $\alpha=0.05$ within a wide range of $s$, but they have inflated type-I error after $s$ passes a transition point. Panels (b)-(d) compare the testing powers under three different alternative hypotheses: $H_a: \bbeta_0^{(1)} - \bbeta_0^{(j)} = \Delta$, where $2\leq j\leq s, \Delta=0.4, 0.6$ and $1$, respectively. We see that the power increases as $N$ increase and $\Delta$ increases. These findings confirm the asymptotic results stated in Theorem \ref{theorem5}.

\begin{figure}[!htb]
\protect\caption{Probability $P(T'_{\calS} > c'_{\calS})$ under different settings of the number of sub-samples and the sample size, using 4 knots.}
\centering \includegraphics[scale=0.6]{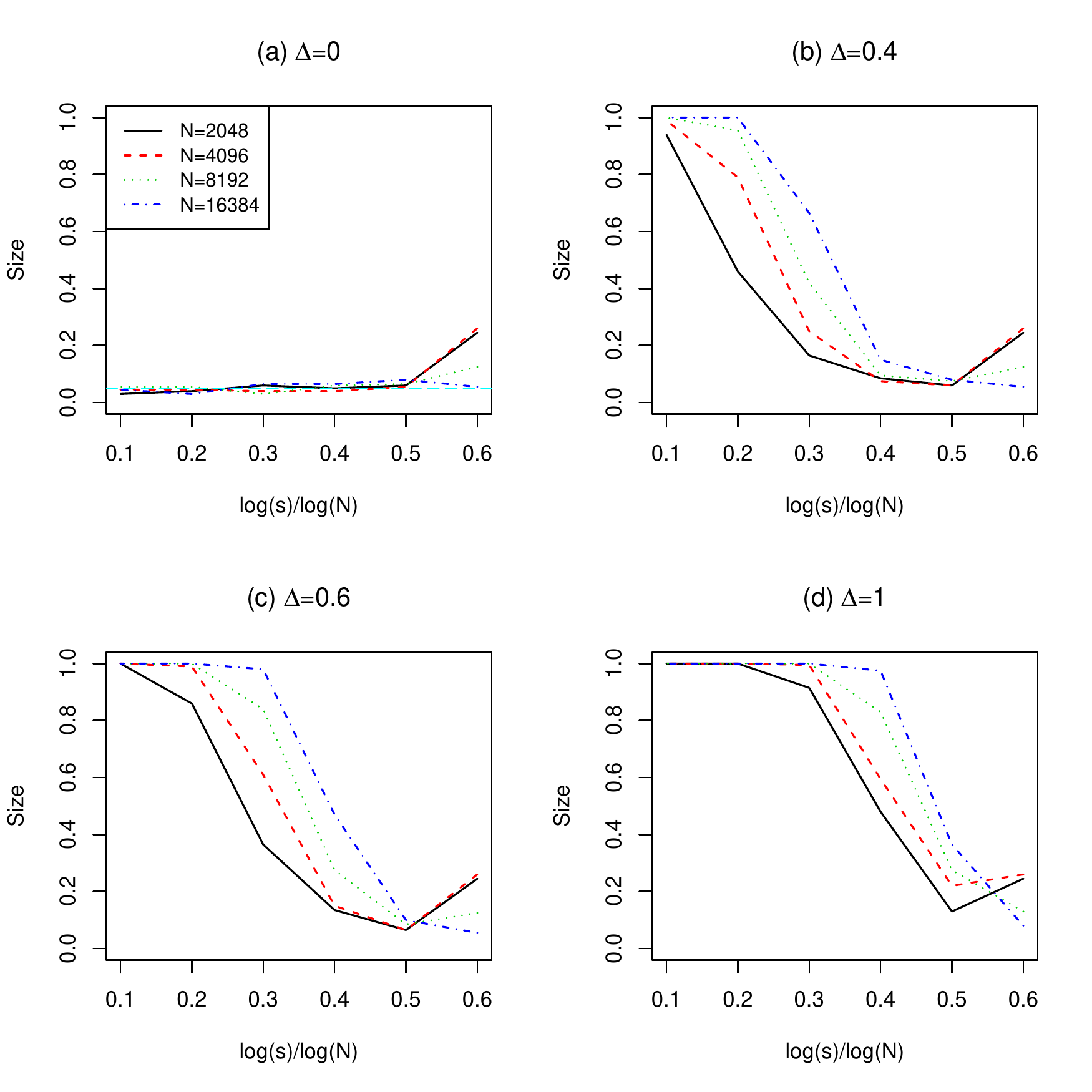}\label{fig:figure6}
\end{figure}

Finally, we evaluate the homogeneity tests for one non-linear function using the test statistic constructed based on Theorem \ref{theorem6}. The null and alternative hypotheses are formulated as follows:
\begin{eqnarray*}
H_0: g_{01}^{(1)}(Z_1) = \cdots = g_{01}^{(s)}(Z_1) \quad {\rm vs} \quad H_a: g_{01}^{(1)}(Z_1) =  g_{01}^{(j)}(Z_1) + \Delta (Z_1+1),j \geq 2,
\end{eqnarray*}
where $\Delta=0, 0.5, 1.0$ and $1.5$. In the simulations, we set $g_{01}^{(j)}(Z_1)=5 \sin \{2\pi(Z_1+1)\}, j\geq 2$ as used previously for the null hypothesis. The results are summarized in Figure \ref{fig:figure7}, where the $y$-axis shows the probability $P(-\frac{1}{3\sigma^2}n \cdot {\rm LRT}_{nk}^s > \chi^2_{u_N,1-\alpha})$ under different settings of $s$ and $N$ and $\chi^2_{u_N,1-\alpha}$ is the $(1-\alpha)$-th quantile of $\chi^2_{u_N}$. From Panel $\Delta=0$ of Figure \ref{fig:figure7}, we see that the proposed test statistic $-\frac{1}{3\sigma^2}n \cdot {\rm LRT}_{nk}^s$ has appropriate type-I error around the nominal level $\alpha=0.05$ if $s$ is not too large nor too small ($\log(s)/\log(N)>q/2$), but they have inflated type-I error after $s$ passes a transition point. As we use a fixed number of knots, the transition point shifts to the right as $N$ increases so that $J_N \ll n^{1/2}$. Panels (b)-(d) compare the testing powers under three different alternative hypotheses with $\Delta=0.5, 1.0$ and $1.5$, respectively. We see that the power increases as $N$ increase and $\Delta$ increases. Similar with observations in \cite{lu2016}, the powers for $\Delta>0$ return back to 1 when $s$ is very large. This can still be explained by highly deviated limiting distribution of the test statistic from both null and alternative hypotheses. Therefore, the homogeneity test does not perform well for very large $s$.

\begin{figure}[!htb]
\protect\caption{Probability $P(-\frac{1}{3\sigma^2}n \cdot {\rm LRT}_{nk}^s > \chi^2_{u_N,1-\alpha})$ under different settings of the number of sub-samples and the sample size, using 4 knots, where $\chi^2_{u_N,1-\alpha}$ is the $(1-\alpha)$-th quantile of $\chi^2_{u_N}$. }
\centering \includegraphics[scale=0.6]{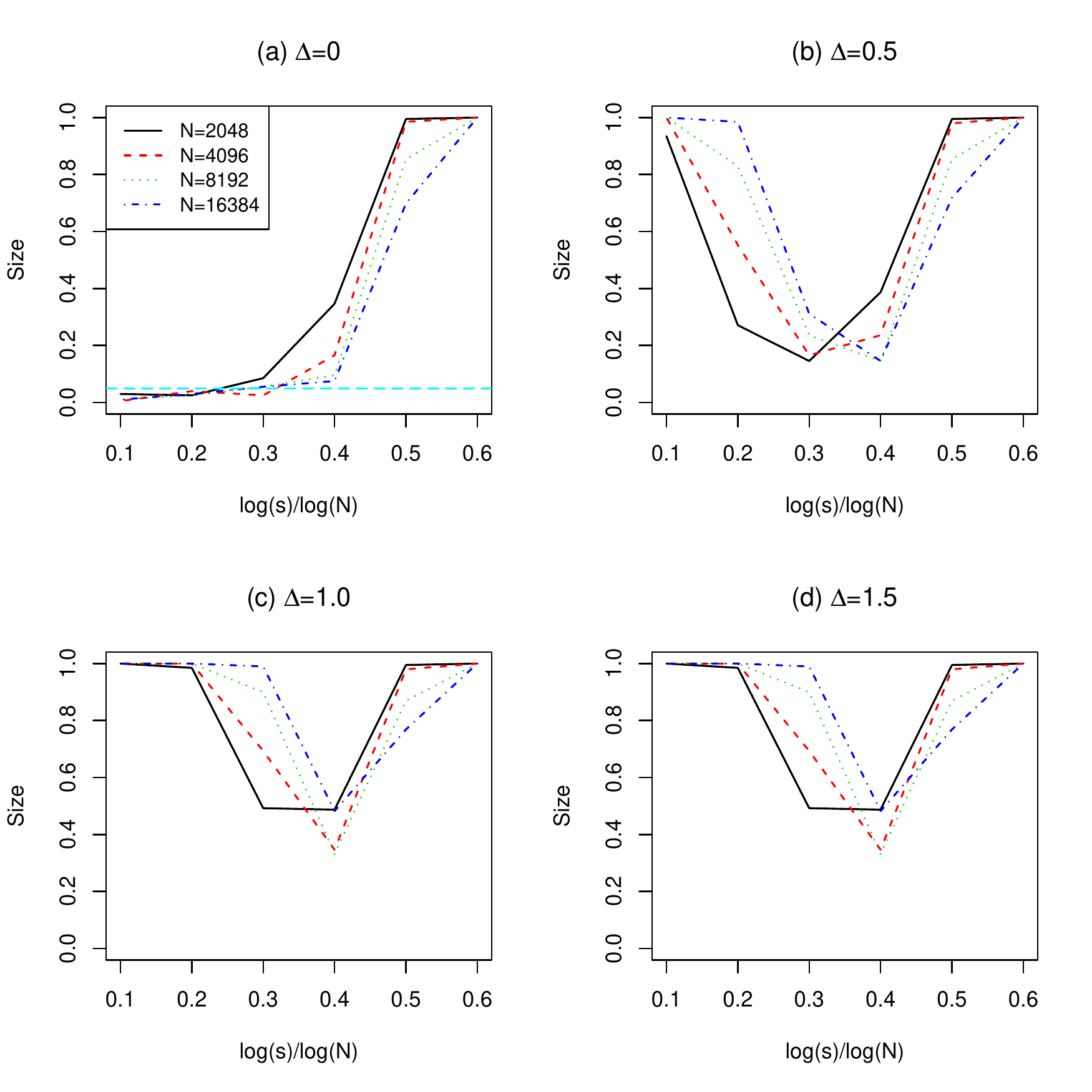}\label{fig:figure7}
\end{figure}

\section{Real data application}

We apply the proposed divide-and-conquer strategy for APLMs to the Medicare Provider Utilization and Payment Data (the Physician and Other Supplier Public Use File), with information on services and procedures provided to Medicare beneficiaries by physicians and other healthcare professionals. This dataset was prepared by the Centers for Medicare \& Medicaid Services (CMS), as part of the Obama Administration efforts to make our healthcare system more transparent, affordable, and accountable. We downloaded the dataset ``Medicare Physician and Other Supplier Data CY 2014" from \url{www.CMS.gov} with more than nine million records for health care providers from the U.S. or U.S. possessions. We focus on the subset consisting of 50 U.S. states and the District of Columbia (DC), which account for the majority part of the whole dataset.

Our goal is to model the outcome variable ``average\_Medicare\_standardized\_amt" (average amount that Medicare paid after beneficiary deductible and coinsurance amounts have been deducted for the line item service and after standardization of the Medicare payment has been applied) on other covariates, including gender or entity of provider, provider type, Medicare participation status, place of service, HCPCS drug indicator, number of distinct Medicare beneficiaries (``bene\_unique\_cnt"), number of services provided (``bene\_day\_srvc\_cnt"), and number of distinct Medicare beneficiary/per day services (``line\_srvc\_cnt"). Detailed explanations of these variables can be found in the official website \url{www.CMS.gov}. All covariates except the last three are categorical variables, and particularly the variable for provider type has 91 categories. Because those three quantitative variables are all count data, we take the $\log_{10}$-transformation and rescale each of them to the range $[-1,1]$ by using the formula $(\boldZ-\min \boldZ)/(\max \boldZ - \min \boldZ) \times 2 -1$. Also, we apply the $\log_{10}$-transformation to the outcome variable, which is skewed to the right. By excluding those records with value 0 for quantitative variables and choosing records with overlapping ranges for last three variables across states, the working dataset has 9,263,068 records, and the corresponding file size is greater than 2GB. It is hard to apply any complicated model fitting with iterative algorithms on a single PC with limited memory.

Therefore, we turn to the developed divide-and-conquer strategy. It is natural to split the data by location, such as states or counties. According to our theoretical results, the number of sub-populations cannot be too large. The number of counties is more than 3,000 in U.S., while $\sqrt{9,263,068} \approx 3044$. Thus, we split the whole dataset by states and DC, resulting in 51 sub-populations. The number of records for each sub-population varies from 14,809 (Alaska) to 719,970 (California), and the median number is 128,069. It is reasonable to hypothesize that those categorical covariates are heterogeneous because their effects on the average amount that Medicare paid may vary across states. On the other hand, the outcome variable is the standardized payment by removing geographic differences in payment rates for individual services, and all three quantitative covariates are numbers of services and beneficiaries. Then it is reasonable to assume the effects of quantitative covariates are homogeneous.

We choose B-splines with degree of 3 to approximate the non-parametric functions of those three quantitative covariates. Assumption (A4) requires that the number of internal knots should be much smaller than $N^{\frac{1}{4}}\approx 55$. Additionally, we expect these curves are smooth. Thus, we set the number of internal knots as 5. Noting that the sizes of sub-populations are different, rather than a simple average to obtain the aggregated curves, a weighted average is employed by using weights $n_j/\sum_{j=1}^s n_j$, where $n_j$ is the size of the $j$th sub-sample.

\begin{figure}[!htb]
\protect\caption{Box-plots of heterogeneous parameters without aggregation of commonality.across 50 states and the DC: the left panel shows estimates of gender/entity, Medicare participation status, place of service and HCPCS drug status; the right panel shows estimates of 90 provider types versus the reference type.}
\centering \includegraphics[scale=0.5]{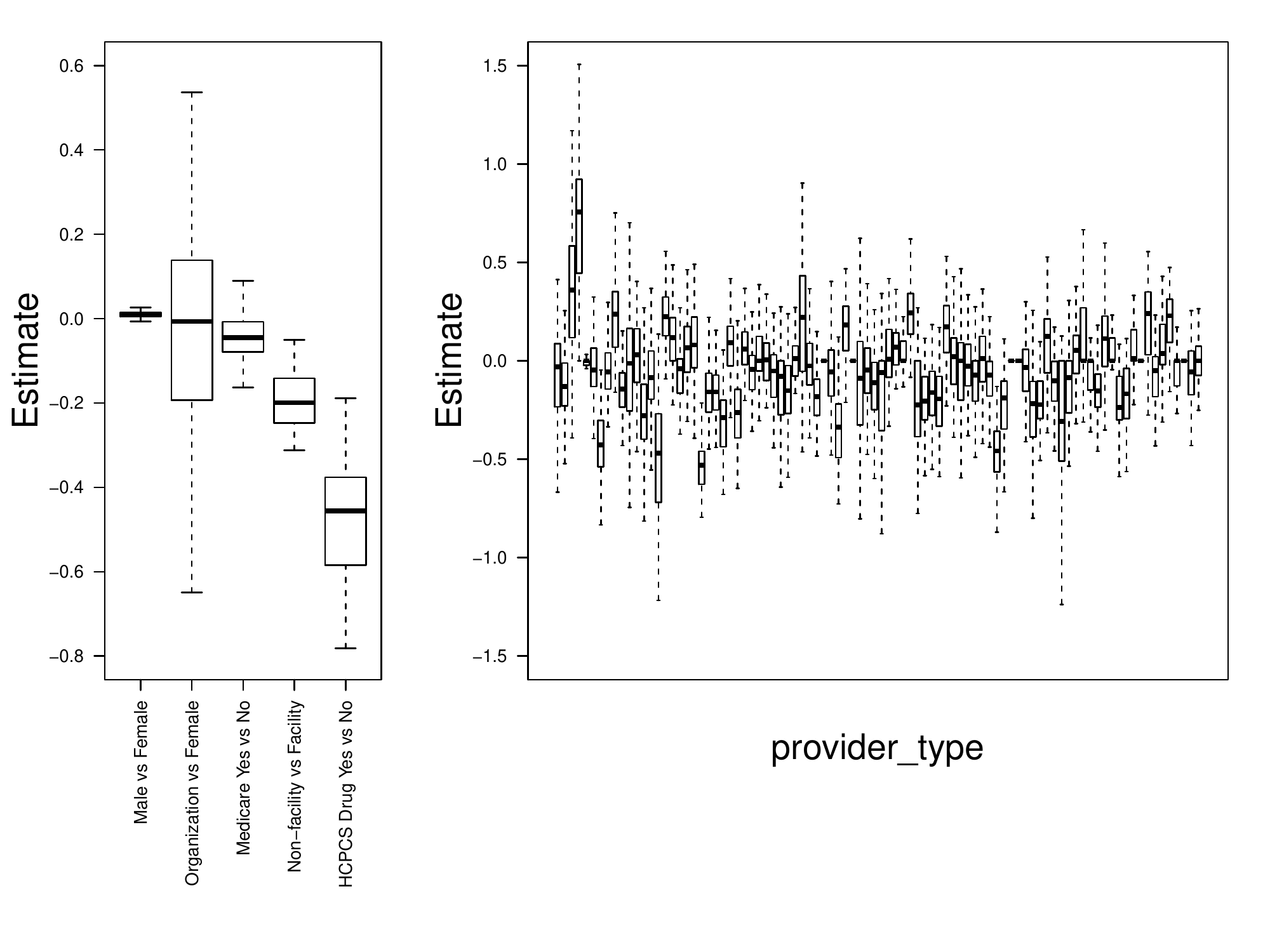}\label{fig:figure8}
\end{figure}

Since the effects of those categorical covariates are allowed to be heterogeneous, we use box-plots to summarize the variabilities of their estimates across 51 sub-populations. From Figure \ref{fig:figure8}, which displays the extent of heterogeneity, we can see that only the effect of male versus female has small degree of heterogeneity around 0, and all the other estimates have substantial variabilities. We further test the heterogeneity of the gender effects across 51 sub-populations via the testing procedure proposed in Section \ref{sec:test_hetero}. The bootstrapped (based on 500 bootstrapping samples) critical value is $c_{\calS}'=2.69$ under $\alpha=1\%$, while the test statistic $W_{\calS}'=10.09$ (p-value is 0). Although the range of this effect seems is small, with such a large sample size, we can easily detect a small heterogeneity across sub-populations. In summary, it implies that the effects of categorical covariates on the average amount that Medicare paid vary a lot across states.

\begin{figure}[!htb]
\protect\caption{Estimates of smooth functions based on each sub-population and the aggregation. (a): the estimated curves for ``bene\_unique\_cnt"; (b): the estimated curves for ``line\_srvc\_cnt"; (c): the estimated curves for ``line\_srvc\_cnt".}
\centering \includegraphics[scale=0.6]{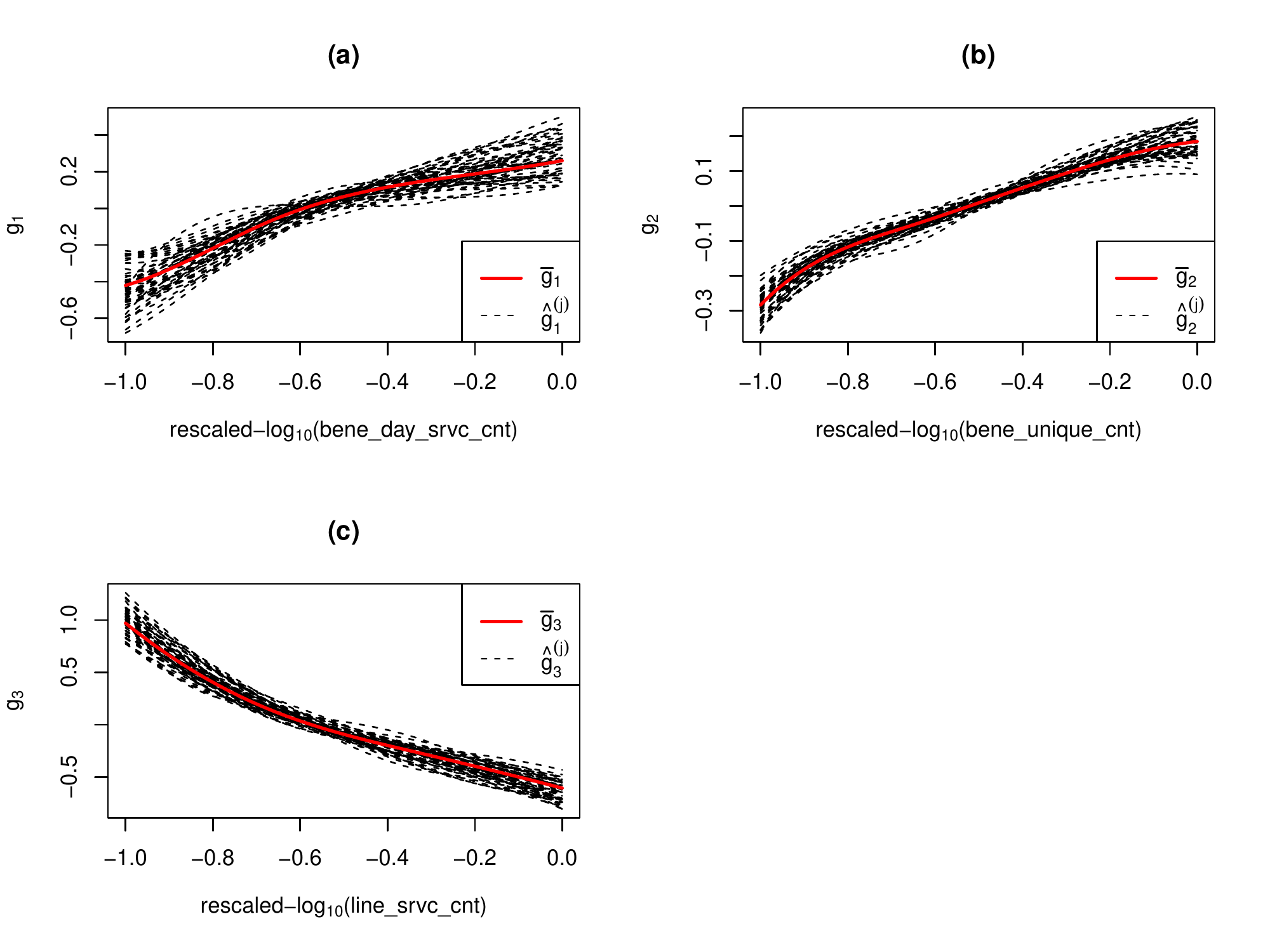}\label{fig:figure9}
\end{figure}

Figure \ref{fig:figure9} presents the non-parametric estimates of the effects of those three quantitative covariates. The largest value of each quantitative covariate is different across states, so we only plot aggregated curves on the common range from -1 to 0. From panels (a)-(c) of the figure, for each covariate, we can see estimated curves from 51 sub-samples (dashed lines in black color) are almost parallel to each other within a narrow band, while the aggregated curve (solid line in red color) is right in the middle of those sub-sample specific curves. In addition, we are interested in the validity of the aggregation for commonality, noting that the sub-popualtion curves at boundaries vary more than they vary around the middle parts. Then we test the homogeneity of the non-linear compoents across 51 sub-populations on the range from $-0.8$ to $-0.1$ via the testing procedure proposed in Section \ref{sec:test_homo}. The test statistic $-\frac{1}{3\sigma^2 }n \cdot {\rm LRT}_{nk}^s=480.24$ and $u_N=500$. Therefore, the resulting $p=0.73$ means failing to reject the null hypothesis, and aggregation for commonality is a valid strategy.
%

\section{Summary}

In this paper, we develop a framework for additive partially linear models for massive heterogeneous data, using the divide-and-conquer strategy. As summarized in \cite{wang2015statistical}, the divide-and-conquer strategy is one of the three commonly used strategies for analyzing massive data, with the other two being the sub-sampling strategy and the sequential updating strategy. However, the sub-sampling and sequential updating strategies are only suitable for analyzing homogeneous massive data. We can combine the divide-and-conquer and sub-sampling strategies to analyze heterogeneous data, by dividing the data into homogeneous subgroups and then conducting sub-sampling within each subgroup. We can also combine the divide-and-conquer and sequential updating strategies to analyze heterogeneous data, by dividing the data into homogeneous subgroups and then conducting sequential updating within each subgroup.

The framework developed in this paper extends the partially linear framework proposed in \cite{Zhao2016}. Their partially linear framework considers only one common feature, using the smoothing-splines technique to fit the non-parametric function based on the general reproducing kernel Hilbert space (RKHS) theory \citep{wahba1990spline}. Although the smoothing-splines technique and the RKHS theory have been well developed in the framework of generalized additive models \cite{hastie1990generalized}, we find it very hard to extend them to our goal of analyzing massive data with multiple common features. Instead, we adopt polynomial splines for modeling the non-parametric effects of multiple common features simultaneously across all sub-populations while exploring heterogeneity of each sub-population. The proposed methods can be implemented easily and perform well in both simulation studies and the real data application.

\appendix

\section*{Appendix}

\setcounter{equation}{0} \setcounter{lemma}{0}
\global\long\def\theequation{A.\arabic{equation}}
 \global\long\def\thesubsection{A.\arabic{subsection}}
 \global\long\def\thelemma{A.\arabic{lemma}}

\subsection*{A.1 Technical lemmas for Section \ref{subsec:ind}}

Define the centered version of B-spline basis as
\begin{eqnarray*}
b^*_{m,k}(z_k)=b_{m,k}(z_k) - \frac{E[b_{m,k}]}{E[b_{1,k}]}b_{1,k}(z_k),\ k=1,\ldots,K, m=-\varrho+1,\ldots,J_N,
\end{eqnarray*}
and the standardized version of B-spline basis as
\begin{eqnarray*}
B_{m,k}(z_k)=\frac{b^*_{m,k}(z_k)}{\|b^*_{m,k} \|_{2k}}, \ m=-\varrho+1,\ldots, J_N, k=1,\ldots,K.
\end{eqnarray*}
Then the minimization problem (\ref{eq:j-popu-equi}) is equivalent to the following minimization problem:
\begin{eqnarray*}
(\wh{\bbeta}^{(j)}, \wh{\bgamma}^{(j)}) = \argmin_{\stgbeta \in {\mathcal R}^d,\ \stggamma \in {\mathcal R}^{K(J_N+\varrho)}} \frac{1}{2} \sum_{i \in G_j} \left[ Y_i - \boldX_i\trans \bbeta -  \boldB(\boldZ_i)\trans \bgamma\right]^2,
\end{eqnarray*}
where $\boldB(\boldz)=\{ B_{m,k}(z_k), m=-\varrho+1,\ldots,J_N, k=1,\ldots,K \}\trans$. Here, to abuse the notation, we still use $\wh{\gamma}^{(j)}$.   Then $\wh{g}^{(j)}(\boldz)=\wh{\bgamma}\trans \boldB(\boldz)$ is a spline estimator of $g_0$ for the $j$th sub-population, and the centered spline estimators of a component function is
\begin{eqnarray*}
\wh{g}_k^{(j)} (z_k) = \sum_{m=-\varrho+1}^{J_N} \wh{\gamma}_{m,k} B_{m,k}(Z_k) -\frac{1}{n} \sum_{i \in G_j} \sum_{m=-\varrho+1}^{J_N} \wh{\gamma}_{m,k} B_{m,k} (Z_{ik}).
\end{eqnarray*}
In practice, basis $\{ b_{m,k}, m=-\varrho+1, \ldots, J_N, k=1,\ldots,K \}$ is used for computational implementation, while $\{ B_{m,k} \}$ is convenient for asymptotic analysis.

\cite{DeBoor1978} showed that for any function $f \in \calH$ and $N \geq 1$, there exists a function $\wt{f} \in \calS_N$ such that $\| \wt{f} - f \|_{\infty} \leq C h_N^p$. Thus, for $g_0$ satisfying Assumption (A1), there exists a $\wt{g}^{(j)}(\boldz)=\boldB\trans(\boldz)\wt{\bgamma}^{j} \in \calG_n^{(j)}$ s.t. $\| \wt{g}^{(j)} - g_0 \|_{\infty}=O(h_N^p)$ and $\wt{g}^{(j)}$(\boldz) is the best least-squares projection of $g_0(\boldz)$ into the space $\calG_n^{(j)}$, implying
\begin{eqnarray} \label{eq:projection}
\langle \wt{g}^{(j)}(\boldz) - g_0(\boldz),  \boldB(\boldz) \rangle_{jn} =0, \ \ j=1,\ldots,s.
\end{eqnarray}
Define
\begin{eqnarray*}
\wt{\bbeta}^{(j)} =\argmin_{\bbeta} \frac{1}{2} \sum_{i \in G_j} \left[ Y_i - \wt{g}^{(j)}(\boldZ_i) - \boldX_i\trans \bbeta \right]^2,
\end{eqnarray*}
and let $m_{0i}^{(j)}\equiv m_0^{(j)}(\boldT_i)=g_0(\boldZ_i)+\boldX_i\trans \bbeta_0^{(j)}$, $\wt{m}_{0}^{(j)}(\bft) =\wt{g}^{(j)}(\boldz)+\bfx\trans \bbeta_0^{(j)}$, and $\wt{m}_{0i}^{(j)}\equiv \wt{m}_0^{(j)}(\boldT_i)=\wt{g}^{(j)}(\boldZ_i)+\boldX_i\trans \bbeta_0^{(j)}$.

Additionally, let $\btheta=\left(\begin{smallmatrix}\bgamma \\ \bbeta\end{smallmatrix}\right), \wh{\btheta}^{(j)} =\left(\begin{smallmatrix}\wh{\bgamma}^{(j)} \\ \wh{\bbeta}^{(j)} \end{smallmatrix} \right), \wt{\btheta}^{(j)}=\left(\begin{smallmatrix}\wt{\bgamma}^{(j)} \\ \wt{\bbeta}^{(j)}\end{smallmatrix}\right), \wh{l}_n^{(j)}(\btheta)=l_n^{(j)}(\bgamma,\bbeta)$, and
$$ \wt{m}_i^{(j)}\equiv \wt{m}^{(j)}(\boldT_i)=\wt{g}^{(j)} + \boldX_i\trans \wt{\bbeta} =\boldB\trans (\boldZ_i)\wt{\bgamma}^{(j)} + \boldX_i\trans \wt{\bbeta}^{(j)}.$$
Define
\begin{eqnarray*}
V_n^{(j)}\doteq\frac{ \partial^2 \wh{l}_n^{(j)}(\btheta) }{\partial \btheta \partial \btheta\trans} =\frac{1}{n}\sum_{i \in G_j} \left\{ \begin{array}{cc}
(\boldB(\boldZ_i))^{\otimes 2} & \boldB(\boldZ_i) \boldX_i\trans \\
\boldX_i \boldB\trans(\boldZ_i) & \boldX_i^{\otimes 2}
\end{array}\right\}.
\end{eqnarray*}

\begin{lemma} \label{lemmaA1}
Under Assumptions (A1)-(A4), for each sub-population $j$, $$\sqrt{n}(\wt{\bbeta}^{(j)}-\bbeta_0^{(j)}) \xrightarrow{d} \calN(\bfzero, \boldA^{-1} \bSigma_1 \boldA^{-1}),$$ where $\boldA=E(\boldX^{\otimes 2})$ and $\bSigma_1 = E(\varepsilon^2 \boldX^{\otimes 2})$.
\end{lemma}

\noindent{\it Proof.} Let $\wt{\bdelta}^{(j)}=\sqrt{n}(\wt{\bbeta}^{(j)}-\bbeta_0^{(j)})$. Then $\wt{\bdelta}^{(j)}$ minimizes
\begin{eqnarray*}
\wt{l}_n^{(j)}(\bdelta) = \frac{1}{2} \sum_{i \in G_j} \left[ \left( Y_i - \wt{m}_{0i}^{(j)} -\frac{1}{\sqrt{n}} \boldX\trans \bdelta \right)^2 - \left( Y_i - \wt{m}_{0i}\right)^2 \right].
\end{eqnarray*}
Let $\boldA_n^{(j)}=\frac{1}{n}\sum_{i \in G_j} \boldX_i^{\otimes 2}$. By taking derivatives with respect to $\bdelta$, we obtain
\begin{eqnarray*}
\frac{\partial \wt{l}_n^{(j)}(\bdelta)}{\partial \bdelta} = \boldA_n^{(j)} \bdelta - \frac{1}{\sqrt{n}} \sum_{i \in G_j} ( Y_i - \wt{m}_{0i}^{(j)})\boldX_i = \bfzero,
\end{eqnarray*}
which implies
$$ \wt{\bdelta}^{(j)} =\frac{1}{\sqrt{n}} (\boldA_n^{(j)})^{-1} \sum_{i \in G_j} \varepsilon_i \boldX_i + \frac{1}{\sqrt{n}} (\boldA_n^{(j)})^{-1} \sum_{i \in G_j} \big(g_0(\boldZ_i) - \wt{g}^{(j)}(\boldZ_i) \big) \boldX_i.
$$
With similar arguments with those of Lemma A.1 in \cite{liu2011} and the fact $\| \wt{g}^{(j)} - g_0 \|_{\infty}=O(h_N^p)$, the lemma follows. \hfill $\Box$

\begin{lemma} \label{lemmaA2}
Under Assumptions (A1)-(A4), if $J_N \ll \frac{n}{(\log n)^2}$, there exists a constant $C$ such that $\sup_{j} \| (V_n^{(j)})^{-1} \|_2 \leq C$, a.s.
\end{lemma}

\noindent{\it Proof.} For each sub-population $j$, Lemma A.2 in \cite{liu2011} showed there exists a constant $C$ such that $\| (V_n^{(j)})^{-1} \|_2 \leq C$, a.s., if
\begin{eqnarray*}
\sup_{f_1, f_2 \in \calG_n^{(j)}} \left| \frac{\langle f_1,f_2\rangle_{jn} - \langle f_1,f_2\rangle}{\|f_1\|_2 \|f_2\|_2} \right| = O \left( \frac{\log n}{(n h_N)^{1/2}} \right)=o(1), \ \ a.s.,
\end{eqnarray*}
by Lemma A.8 in \cite{Wang2007}. Here constant $C$ is taken to be large enough to ensure that the above result holds for all $j=1, \cdots, s$.
The condition $J_N \ll \frac{n}{(\log n)^2}$ implies $O \left( \log n/(n h_N)^{1/2} \right)=o(1)$. Therefore, the lemma is proved. \hfill $\Box$

\begin{lemma} \label{lemmaA3}
Under Assumptions (A1)-(A4), for each sub-population $j$, we have
$$\left\| \wh{\btheta}^{(j)} - \wt{\btheta}^{(j)} \right\| = O_P\left( J_N^{1/2} n^{-1/2} +h_N^p\right).$$
\end{lemma}

\noindent{\it Proof.} It follows that
\begin{eqnarray*}
\left.\frac{ \partial \wh{l}_n^{(j)}(\btheta)}{\partial \btheta}\right|_{\btheta = \wh{\btheta}^{(j)}} - \left.\frac{ \partial \wh{l}_n^{(j)}(\btheta) }{\partial \btheta}\right|_{\btheta = \wt{\btheta}^{(j)}} = \left.\frac{ \partial^2 \wh{l}_n^{(j)}(\btheta) }{\partial \btheta \partial \btheta\trans}\right|_{\btheta = \overline{\btheta}^{(j)}} \big( \wh{\btheta}^{(j)} - \wt{\btheta}^{(j)} \big),
\end{eqnarray*}
where $\overline{\btheta}^{(j)}$ is between $\wh{\btheta}^{(j)}$ and $\wt{\btheta}^{(j)}$. Thus, we have
\begin{eqnarray*}
\wh{\btheta}^{(j)} - \wt{\btheta}^{(j)} = -\left(  \left.\frac{ \partial^2 \wh{l}_n^{(j)}(\btheta) }{\partial \btheta \partial \btheta\trans}\right|_{\btheta = \overline{\btheta}^{(j)}} \right)^{-1} \left.\frac{ \partial \wh{l}_n^{(j)}(\btheta) }{\partial \btheta}\right|_{\btheta = \wt{\btheta}^{(j)}}.
\end{eqnarray*}
We can write
\begin{eqnarray*}
\frac{1}{n} \frac{\partial \wh{l}_n^{(j)} (\btheta)}{\partial \btheta} \Big|_{\btheta=\wt{\btheta}^{(j)}} &=&
-\frac{1}{n}\sum_{i\in G_j}(Y_i - m_{0i}^{(j)})\left(\begin{smallmatrix}\boldB(\boldZ_i) \\ \boldX_i\end{smallmatrix}\right)\\
&& + \frac{1}{n}\sum_{i\in G_j}(\wt{g}^{(j)}(\boldZ_i)-g_0(\boldZ_i))\left(\begin{smallmatrix}\boldB(\boldZ_i) \\ \boldX_i\end{smallmatrix}\right) \\
&& + \frac{1}{n}\sum_{i\in G_j} \boldX_i\trans (\wt{\bbeta}^{(j)}-\bbeta_0^{(j)}) \left(\begin{smallmatrix}\boldB(\boldZ_i) \\ \boldX_i\end{smallmatrix}\right).
\end{eqnarray*}
First, by \eqref{eq:projection}, we have $\sum_{i\in G_j}(\wt{g}^{(j)}(\boldZ_i)-g_0(\boldZ_i)) \boldB(\boldZ_i)=\bfzero$. With similar arguments with those of Lemma A.3 in \cite{liu2011}, we have
\begin{eqnarray*}
\left\| \frac{1}{n}\sum_{i\in G_j}(Y_i - m_{0i}^{(j)}) \boldB(\boldZ_i) \right\| &=& O_P\left(  J_N^{1/2} n^{-1/2} \right), \\
\left\| \frac{1}{n}\sum_{i\in G_j}(Y_i - m_{0i}^{(j)}) \boldX_i \right\| &=& O_P\left(  n^{-1/2} \right), \\
\left\| \frac{1}{n}\sum_{i\in G_j}(\wt{g}^{(j)}(\boldZ_i)-g_0(\boldZ_i)) \boldX_i \right\| &=& O_P\left(  h_N^p \right), \\
\left\| \frac{1}{n}\sum_{i\in G_j} \boldX_i\trans (\wt{\bbeta}^{(j)}-\bbeta_0^{(j)}) \boldB(\boldZ_i) \right\| &=&  o_P\left(  J_N^{1/2} n^{-1/2} \right), \\
\left\| \frac{1}{n}\sum_{i\in G_j} \boldX_i\trans (\wt{\bbeta}^{(j)}-\bbeta_0^{(j)}) \boldX_i \right\| &=&  o_P\left( n^{-1/2} \right).
\end{eqnarray*}

Therefore, by Lemma \ref{lemmaA2}, we have
$$ \| \wh{\btheta}^{(j)} - \wt{\btheta}^{(j)} \| \leq \| ( V_n^{(j)})^{-1} \|_2 \left\| \frac{1}{n} \frac{\partial \wh{l}_n^{(j)} (\btheta)}{\partial \btheta} \Big|_{\btheta=\wt{\btheta}^{(j)}} \right\| = O_P\left(  J_N^{1/2} n^{-1/2} + h_N^p\right). $$
\hfill $\Box$

\begin{lemma} \label{lemmaA4}
Under Assumptions (A1)-(A4), for each sub-population $j$, if $J_N \ll \frac{n}{(\log n)^2}$, we have
\begin{eqnarray*}
&&\left\| \wh{g}^{(j)} - g_0 \right\|_2 = O_P\left( J_N^{1/2} n^{-1/2} + h_N^p \right), \\
&&\left\| \wh{g}^{(j)} - g_0 \right\|_{jn} = O_P\left( J_N^{1/2} n^{-1/2} + h_N^p  \right), \\
&&\left\| \wh{g}^{(j)}_k - g_{0k} \right\|_{2k} = O_P\left( J_N^{1/2} n^{-1/2} + h_N^p  \right), \\
&&\left\| \wh{g}^{(j)}_k - g_{0k} \right\|_{jnk} = O_P\left( J_N^{1/2} n^{-1/2} + h_N^p  \right),
\end{eqnarray*}
where $k=1,\ldots,K$.
\end{lemma}

\noindent{\it Proof.} The proof is similar with that of Lemma A.4 in \cite{liu2011} by applying Lemmas \ref{lemmaA2} and \ref{lemmaA3} and noting that
\begin{eqnarray*}
\sup_{f \in \calS_N}  \frac{\| f \|_{jnk}}{\| f \|_{2k}}  = O_P \left( \frac{\log n}{(n h_N)^{1/2}} \right)=o_P(1), \ \ k=1,\ldots,K,
\end{eqnarray*}
which is implied by Lemma A.8 in \cite{Wang2007} under condition $J_N \ll n/(\log n)^2$. \hfill $\Box$

\begin{lemma} \label{lemmaA5}
Under Assumptions (A1)-(A4), for each sub-population $j$, if $n \gg J_N^2$, we have
\begin{eqnarray*}
\frac{1}{n} \sum_{i \in G_j} \wt{\boldX}_i \Gamma (\boldZ_i)\trans \big(\wh{\bbeta}^{(j)}-\bbeta_0^{(j)}\big) &=& o_P(n^{-1/2}), \\
\frac{1}{n} \sum_{i \in G_j} \big(\wh{g}^{(j)}(\boldZ_i)- g_0(\boldZ_i) \big) \wt{\boldX}_i  &=& o_P(n^{-1/2}),
\end{eqnarray*}
\end{lemma}

\noindent{\it Proof.} The proof is similar with that of Lemma A.5 in \cite{liu2011} by making following revisions. We only show the second equality, and the first one can be proved similarly.

Let $w_1(\boldZ,g)=g(\boldZ)\wt{\boldX}$, and it follows
$$ E\big\| w_1(\boldZ,\wh{g}^{(j)}) - w_1(\boldZ,g_0) \big\|^2 =E \left\| (\wh{g}^{(j)}(\boldZ_i)-g_0(\boldZ_i)) \wt{\boldX}_i \right\|^2 \leq O\big( E\| \wh{g}^{(j)} -g_0 \|_2^2 \big). $$
By Lemma A.2 of \cite{Huang1999}, the logarithm of the $\varepsilon$-bracketing number of the class of functions $\calA_1^{(j)}(\delta)=\{ w_1(\cdot, \wh{g}) - w_1(\cdot,g_0) : \wh{g} \in \calG_n^{(j)}, \| \wh{g}-g_0 \|_2 \leq \delta \}$ is $c\{ (J_N - \varrho) \log (\delta/\varepsilon) + \log(\delta^{-1}) \}$. Thus, the corresponding entropy integral $J_{[]}(\delta, \calA_1^{(j)}(\delta), \| \cdot \|_2 ) \leq c \delta \{ (J_N - \varrho)^{1/2} + \log^{1/2} (\delta^{-1}) \}$. According to Lemma 7 of \cite{Stone1986} and Lemma \ref{lemmaA4}, $\| \wh{g}^{(j)} - g_0 \|_{\infty} \leq c J_N^{1/2}\| \wh{g}^{(j)} - g_0 \|_{2}=O_P(J_N n^{-1/2}+ J_N^{1/2}h_N^p)$. Let $r_{n,N}^{-1}=J_N^{1/2} n^{-1/2} +h_N^p$, then
\begin{eqnarray*}
&& E\left| \frac{1}{n} \sum_{i \in G_j} \left\{ \wh{g}^{(j)}(\boldZ_i) - g_0(\boldZ_i) \right\} \wt{\boldX}_i - E \left\{ \wh{g}^{(j)}(\boldZ_i) - g_0(\boldZ_i) \right\} \wt{\boldX}_i\right| \\
&\leq& n^{-1/2} C r_{n,N}^{-1} \left\{ (J_N +\varrho)^{1/2} + \log^{1/2}(r_{n,N})\right\} \\
&& \hspace{1cm} \times
\left[ 1 + \frac{c r_{n,N}^{-1} \left\{ (J_N+ \varrho)^{1/2} + \log^{1/2}(r_{n,N})\right\}}{r_{n,N}^{-2}\sqrt{n}} C_0 \right]\\
&\leq& O(1) n^{-1/2} C r_{n,N}^{-1} \left\{ (J_N +\varrho)^{1/2} + \log^{1/2}(r_{n,N})\right\},
\end{eqnarray*}
where the second inequality is based on the fact $r_{n,N}J_N^{1/2}/\sqrt{n}=O(1)$.

Under condition that $n \gg J_N^2$, we have $J_N/\sqrt{n}=o(1)$, implying $J_N n^{-1/2} + J_N^{1/2} h_N^p =o(1)$, and then $r_{n,N}^{-1}J_N^{1/2}=o(1)$. Therefore, the above term is bounded by $o(n^{-1/2})$. \hfill $\Box$

\subsection*{A.2 Technical lemmas for Section \ref{subsec:agg}}

Let $\wt{g} =\frac{1}{s} \sum_{j=1}^s \wt{g}^{(j)}$. In order to ensure that $\wt{g}\in \calG_N$,  we re-center the individual estimator $\wt{g}^{(j)}(\boldz)$ via $\wt{g}^{(j)} (\boldz) = \sum_{k=1}^K \sum_{m=-\varrho}^{J_N} \wt{\gamma}_{m,k} b_{m,k}(z_k) -\frac{1}{N} \sum_{k=1}^K \sum_{i=1}^N \sum_{m=-\varrho}^{J_N} \wt{\gamma}_{m,k} b_{m,k} (z_{ik})$. To abuse the notation, we still denote the centered estimator as $\wt{g}^{(j)}(\boldz)$. Lemma \ref{lemmaA2} shows $\|(V_n^{(j)})^{-1} \|_2 \leq C$ a.s., $j=1,\ldots,s$, if $n \gg J_N^2$. This property plays a key role in all following proofs. Define $\boldu_i^{(j)}=(V_n^{(j)})^{-1} \left(\begin{smallmatrix}\boldB\big(\boldZ_i\big) \\ \boldX_i\end{smallmatrix}\right)=\{ u_{im}^{(j)}, m=1,\ldots,K(J_N+\varrho)+d \}\trans$, and it follows $\sum_{i \in G_j}(\boldu_i^{(j)})^{\otimes 2} = n (V_n^{(j)})^{-1} $. Let $\bolde_m$ denote a $(K(J_N+\varrho)+d)$-dim vector with its $m$th entry as 1 and 0 otherwise, and thus $ u_{im}^{(j)} =\bolde_m\trans \boldu_i^{(j)}$.

\begin{lemma} \label{lemmaA6}
Under Assumptions (A1)-(A4), if $n \gg J_N^2$, we have
$$\left\| \frac{1}{N} \sum_{j=1}^s \sum_{i\in G_j} \varepsilon_i \boldu_i^{(j)} \right\| =O_P \big(J_N^{1/2} N^{-1/2} \big).$$
\end{lemma}

\noindent{\it Proof.} If follows
$$ \left\| \frac{1}{N} \sum_{j=1}^s \sum_{i\in G_j} \varepsilon_i \boldu_i^{(j)} \right\|^2 = \frac{1}{N^2} \sum_{m=1}^{K(J_N+\varrho)+d} \left\{
\sum_{j=1}^s \sum_{i\in G_j} \varepsilon_i u_{im}^{(j)} \right\}^2. $$
Observing that
\begin{eqnarray*}
\frac{1}{N}E \left\{ \sum_{j=1}^s \sum_{i\in G_j} \varepsilon_i u_{im}^{(j)} \right\}^2 &=& \frac{1}{N^2} E \left[ \sum_{j=1}^s \sum_{i\in G_j}\varepsilon_i^2 ( u_{im}^{(j)})^2 \right] = \frac{\sigma^2}{N^2} E \left[ \sum_{j=1}^s \sum_{i\in G_j} ( u_{im}^{(j)})^2 \right] \\
&=& \frac{\sigma^2}{N^2} E \left[ \sum_{j=1}^s n \bolde_m\trans (V_n^{(j)})^{-1} \bolde_m \right] \leq \frac{N C \sigma^2}{N^2},
\end{eqnarray*}
where the last inequality is due to the fact $\bolde_m\trans (V_n^{(j)})^{-1} \bolde_m \leq \|(V_n^{(j)})^{-1}\|_2^2 \| \bolde_m \|^2 \leq C$ a.s.. Thus
$$\left\| \frac{1}{N} \sum_{j=1}^s \sum_{i\in G_j} \varepsilon_i \boldu_i^{(j)} \right\| =O_P \big(J_N^{1/2} N^{-1/2} \big). $$
\hfill $\Box$

\begin{lemma} \label{lemmaA7}
Under Assumptions (A1)-(A4), if $n \gg J_N^2$, we have
$$ \left\|\frac{1}{N} \sum_{j=1}^s \sum_{i\in G_j} (\wt{g}^{(j)}(\boldZ_i)-g_0(\boldZ_i))\boldu_i^{(j)} \right\| =O_P \big(J_N^{1/2} h_N^{p} \big).$$
\end{lemma}

\noindent{\it Proof.} Note that
\begin{eqnarray*}
\frac{1}{N} \sum_{j=1}^s \sum_{i\in G_j} (\wt{g}^{(j)}(\boldZ_i)-g_0(\boldZ_i))\boldu_i^{(j)}
&=& \frac{1}{N}   \sum_{j=1}^s (V_n^{(j)})^{-1}  \sum_{i\in G_j} (\wt{g}^{(j)}(\boldZ_i)-g_0(\boldZ_i)) \left(\begin{smallmatrix}\boldB\big(\boldZ_i\big) \\ \boldX_i\end{smallmatrix}\right) \\
&=& \frac{1}{N}   \sum_{j=1}^s (V_n^{(j)})^{-1}  \sum_{i\in G_j} (\wt{g}^{(j)}(\boldZ_i)-g_0(\boldZ_i)) \left(\begin{smallmatrix} \bfzero \\ \boldX_i\end{smallmatrix}\right),
\end{eqnarray*}
where the last equality follows from \eqref{eq:projection}.

Therefore, it follows
\begin{eqnarray*}
\left\|\frac{1}{N} \sum_{j=1}^s \sum_{i\in G_j} (\wt{g}^{(j)}(\boldZ_i)-g_0(\boldZ_i))\boldu_i^{(j)} \right\|
\leq   \frac{1}{N}   \sum_{j=1}^s \left\| (V_n^{(j)})^{-1} \right\|_2  \sum_{i\in G_j} \left\| (\wt{g}^{(j)}(\boldZ_i)-g_0(\boldZ_i)) \left(\begin{smallmatrix}\bfzero \\ \boldX_i\end{smallmatrix}\right) \right\|  \\
\leq \frac{C}{N}   \sum_{j=1}^s  \sum_{i\in G_j} \left\| (\wt{g}^{(j)}(\boldZ_i)-g_0(\boldZ_i)) \right\|_{\infty} \left\| \left(\begin{smallmatrix}\bfzero \\ \boldX_i\end{smallmatrix}\right) \right\|_1 = O_{P}(h_N^p).
\end{eqnarray*}
\hfill $\Box$

\begin{lemma} \label{lemmaA8}
Under Assumptions (A1)-(A4), if $n \gg J_N^2$, we have
$$ \left\| \frac{1}{N} \sum_{j=1}^s \sum_{i\in G_j} \boldX_i\trans (\wt{\bbeta}^{(j)}-\bbeta_0^{(j)}) \boldu_i^{(j)}  \right\| =o_P \left( J_N^{1/2} N^{-1/2} \right). $$
\end{lemma}

\noindent{\it Proof.} It follows
$$  \left\| \frac{1}{N} \sum_{j=1}^s \sum_{i\in G_j} \boldX_i\trans (\wt{\bbeta}^{(j)}-\bbeta_0^{(j)}) \boldu_i^{(j)}  \right\|^2
 = \frac{1}{N^2} \sum_{m=1}^{K(J_N+\varrho)+d} \left\{
\sum_{j=1}^s \sum_{i\in G_j} \boldX_i\trans (\wt{\bbeta}^{(j)}-\bbeta_0^{(j)}) u_{im}^{(j)} \right\}^2.$$
The proof of Lemma \ref{lemmaA1} shows
$$ \wt{\bbeta}^{(j)} -\bbeta_0^{(j)} = \frac{1}{n} (\boldA_n^{(j)})^{-1} \sum_{i \in G_j} \varepsilon_i \boldX_i + \frac{1}{n} (\boldA_n^{(j)})^{-1} \sum_{i \in G_j} (\wt{g}^{(j)}(\boldZ_i)-g_0(\boldZ_i)) \boldX_i.$$
Then
\begin{eqnarray*}
&&\sum_{i\in G_j} \boldX_i\trans (\wt{\bbeta}^{(j)}-\bbeta_0^{(j)}) u_{im}^{(j)} \\
&=& \sum_{i_1\in G_j}u_{i_1 m}^{(j)} \boldX_{i_1}\trans \left( \frac{1}{n} (\boldA_n^{(j)})^{-1} \sum_{i_2 \in G_j} \varepsilon_{i_2} \boldX_{i_2} + \frac{1}{n} (\boldA_n^{(j)})^{-1} \sum_{i_2 \in G_j} (\wt{g}^{(j)}(\boldZ_{i_2})-g_0(\boldZ_{i_2})) \boldX_{i_2}\right) \\
&=& \frac{1}{n}  \sum_{i_2 \in G_j} \varepsilon_{i_2} \boldX_{i_2}\trans \left(\sum_{i_1\in G_j} u_{i_1 m}^{(j)} (\boldA_n^{(j)})^{-1}  \boldX_{i_1} \right) \\
&& + \frac{1}{n}  \sum_{i_2 \in G_j} (\wt{g}^{(j)}(\boldZ_{i_2})-g_0(\boldZ_{i_2})) \boldX_{i_2}\trans \left(\sum_{i_1\in G_j} u_{i_1 m}^{(j)} (\boldA_n^{(j)})^{-1}  \boldX_{i_1} \right) \\
&=& \sum_{i \in G_j} \varepsilon_{i} \boldX_{i}\trans \boldv_m^{(j)} +  \sum_{i \in G_j}(\wt{g}^{(j)}(\boldZ_{i})-g_0(\boldZ_{i}))\boldX_{i}\trans \boldv_m^{(j)},
\end{eqnarray*}
where $\boldv_m^{(j)} =\sum_{i\in G_j} u_{i m}^{(j)} (\boldA_n^{(j)})^{-1}  \boldX_{i}$. Thus, we have
\begin{eqnarray*}
&&\frac{1}{N^2} E \left\{ \sum_{j=1}^s \sum_{i\in G_j} \boldX_i\trans (\wt{\bbeta}^{(j)}-\bbeta_0^{(j)}) u_{im}^{(j)} \right\}^2 \\
&=& \frac{1}{N^2} E \left\{ \sum_{j=1}^s \sum_{i\in G_j} \varepsilon_{i} \boldX_{i}\trans \boldv_m^{(j)} +  \sum_{j=1}^s \sum_{i\in G_j}(\wt{g}^{(j)}(\boldZ_{i})-g_0(\boldZ_{i}))\boldX_{i}\trans \boldv_m^{(j)} \right\}^2 \\
&=&\frac{1}{N^2}  E \left\{ \sum_{j=1}^s \sum_{i\in G_j} \varepsilon_{i} \boldX_{i}\trans \boldv_m^{(j)} \right\}^2 + \frac{1}{N^2}  E \left\{ \sum_{j=1}^s \sum_{i\in G_j} (\wt{g}^{(j)}(\boldZ_{i})-g_0(\boldZ_{i}))\boldX_{i}\trans \boldv_m^{(j)} \right\}^2 \\
&\leq&\frac{\sigma^2}{N^2}  E  \sum_{j=1}^s \sum_{i\in G_j} \left\{ \boldX_{i}\trans \boldv_m^{(j)} \right\}^2 + \frac{1}{N} \|\wt{g}^{(j)} (\boldZ_{i})-g_0(\boldZ_{i})\|_{\infty}^2  E  \sum_{j=1}^s \sum_{i\in G_j} \left\{ \boldX_{i}\trans \boldv_m^{(j)} \right\}^2. \\
\end{eqnarray*}

Observing that
\begin{eqnarray*}
\sum_{i\in G_j} \left\{ \boldX_{i}\trans \boldv_m^{(j)} \right\}^2 &=&  \sum_{i\in G_j} (\boldv_m^{(j)})\trans \boldX_i \boldX_{i}\trans \boldv_m^{(j)} = n (\boldv_m^{(j)})\trans \boldA_n^{(j)} \boldv_m^{(j)} \\
&=& \frac{1}{n} \left( \sum_{i\in G_j} u_{im}^{(j)} \boldX_i\trans (\boldA_n^{(j)})^{-1} \right)\boldA_n^{(j)} \left( \sum_{i\in G_j} u_{im}^{(j)} \boldA_n^{(j)})^{-1} \boldX_i \right) \\
&=& \left( \frac{1}{\sqrt{n}}  \sum_{i\in G_j} u_{im}^{(j)} \boldX_i\right)\trans (\boldA_n^{(j)})^{-1}  \left( \frac{1}{\sqrt{n}}  \sum_{i\in G_j} u_{im}^{(j)} \boldX_i\right)
\end{eqnarray*}
Based on the Central Limit Theorem and Slutsky's Theorem, it follows
$$E \left\{\sum_{i\in G_j} \left( \boldX_{i}\trans \boldv_m^{(j)} \right)^2 \right\}=O(1).$$
Therefore,
\begin{eqnarray*}
\frac{1}{N^2} E \left\{ \sum_{j=1}^s \sum_{i\in G_j} \boldX_i\trans (\wt{\bbeta}^{(j)}-\bbeta_0^{(j)}) u_{im}^{(j)} \right\}^2 =O\left(\frac{s}{N^2} +\frac{s h_N^{2p}}{N}\right)=o\left( \frac{1}{N} \right),
\end{eqnarray*}
noting that $h_N^{2p} \ll N^{-1}$. Then, we have
$$ \left\| \frac{1}{N} \sum_{j=1}^s \sum_{i\in G_j} \boldX_i\trans (\wt{\bbeta}^{(j)}-\bbeta_0^{(j)}) \boldu_i^{(j)}  \right\| =o_P \left( J_N^{1/2} N^{-1/2} \right). $$
\hfill $\Box$

\subsection*{A.3 Proofs of theorems}

\noindent{\it Proof of Theorem \ref{theorem1}}. The results about $\wh g^{(j)}$ are implied by Lemma \ref{lemmaA4} directly. We only need to prove the stated result about $\wh\bbeta^{(j)}$. Note that the ondition that $J_N^2 \ll n$ implies that $J_N \ll \frac{n}{(\log n)^2}$. Also the condition that $J_N \gg n^{1/(2p)}$ implies that $h_N^p=O(J_N^{-p}) \ll n^{-1/2}$. Therefore, the stated result about $\wh\bbeta^{(j)}$ can be showed by Lemmas \ref{lemmaA1}-\ref{lemmaA5}, following the same argument of proving Theorem 1 in \cite{liu2011}.

\noindent{\it Proof of Theorem \ref{theorem2}}. We first quantify $\|\wh{g}^{(j)} - g_0 \|_2 $. Noting  $\left\|\wt{g} - g_0 \right\|_2 \leq \left\|\wt{g} - g_0 \right\|_{\infty}=O(h_N^p)$ and
\begin{eqnarray*}
\frac{1}{s} \sum_{j=1}^s \left(\wh{g}^{(j)}(\boldz) -\wt{g}^{(j)}(\boldz)\right) =\frac{1}{s} \boldB\trans(\boldz) \sum_{j=1}^s (\wh{\bgamma}^{(j)} -\wt{\bgamma}^{(j)} ),
\end{eqnarray*}
we have
\begin{eqnarray*}
&& \left\| \frac{1}{s} \sum_{j=1}^s \left(\wh{g}^{(j)}(\boldz) -\wt{g}^{(j)}(\boldz)\right) \right\|_2^2 \\
&=& \int_{[0,1]^K} \left[ \frac{1}{s} \sum_{j=1}^s \left(\wh{g}^{(j)}(\boldz) -\wt{g}^{(j)}(\boldz)\right) \right]^2 f(\boldz) d\boldz \\
&=&\frac{1}{s} \sum_{j=1}^s \left(\wh{\bgamma}^{(j)} -\wt{\bgamma}^{(j)} \right)\trans \left[E \boldB(\boldz) \boldB\trans(\boldz)\right]\sum_{j=1}^s \left(\wh{\bgamma}^{(j)} -\wt{\bgamma}^{(j)} \right) \\
&\leq& \frac{C}{s^2} \left\| \sum_{j=1}^s \left(\wh{\bgamma}^{(j)} -\wt{\bgamma}^{(j)} \right) \right\|^2 \leq \frac{C}{s^2} \left\| \sum_{j=1}^s \left(\wh{\btheta}^{(j)} -\wt{\btheta}^{(j)} \right) \right\|^2.
\end{eqnarray*}
Then we consider $ \left\|\frac{1}{s} \sum_{j=1}^s \left(\wh{\btheta}^{(j)} -\wt{\btheta}^{(j)} \right) \right\|$. The proof of Lemma \ref{lemmaA3} implies that  $$\wh{\btheta}^{(j)} -\wt{\btheta}^{(j)}=(V_n^{(j)})^{-1} \frac{1}{n} \frac{\partial \wh{l}_n^{(j)} (\btheta)}{\partial \btheta} \Big|_{\btheta=\wt{\btheta}^{(j)}},$$
where $\frac{1}{n} \frac{\partial \wh{l}_n^{(j)} (\btheta)}{\partial \btheta} \Big|_{\btheta=\wt{\btheta}^{(j)}}$ is equal to
\begin{eqnarray*}
\left\{-\frac{1}{n}\sum_{i\in G_j}(Y_i - m_{0i}^{(j)})
+ \frac{1}{n}\sum_{i\in G_j}(\wt{g}^{(j)}(\boldZ_i)-g_0(\boldZ_i))
+ \frac{1}{n}\sum_{i\in G_j} \boldX_i\trans (\wt{\bbeta}^{(j)}-\bbeta_0^{(j)})\right\} \left(\begin{smallmatrix}\boldB(\boldZ_i) \\ \boldX_i\end{smallmatrix}\right).
\end{eqnarray*}
Then $\frac{1}{s} \sum_{j=1}^s \left(\wh{\btheta}^{(j)} -\wt{\btheta}^{(j)} \right)$ is equal to
\begin{eqnarray*}
-\frac{1}{N} \sum_{j=1}^s \sum_{i\in G_j} \varepsilon_i \boldu_i^{(j)}
+ \frac{1}{N} \sum_{j=1}^s \sum_{i\in G_j} (\wt{g}^{(j)}(\boldZ_i)-g_0(\boldZ_i))\boldu_i^{(j)}
+ \frac{1}{N} \sum_{j=1}^s \sum_{i\in G_j} \boldX_i\trans (\wt{\bbeta}^{(j)}-\bbeta_0^{(j)}) \boldu_i^{(j)}.
\end{eqnarray*}
Therefore, combining Lemmas \ref{lemmaA6}-\ref{lemmaA8}, we have
$$ \frac{1}{s}\left\| \sum_{j=1}^s \left(\wh{\btheta}^{(j)} -\wt{\btheta}^{(j)} \right) \right\| = O_P \left( J_N^{1/2} N^{-1/2} + h_N^p\right).$$
and further we have,
\begin{eqnarray*}
\| \overline{g} - g_0 \|_2 &=& \left\| \frac{1}{s} \sum_{j=1}^s \wh{g}^{(j)} -\frac{1}{s} \sum_{j=1}^s \wt{g}^{(j)} + \wt{g} - g_0 \right\|_2 \leq \left\| \frac{1}{s} \sum_{j=1}^s \left(\wh{g}^{(j)}-\wt{g}^{(j)}\right) \right\|_2 + \left\|\wt{g} - g_0 \right\|_2 \\
&=& O_P \left( J_N^{1/2} N^{-1/2} + h_N^p\right).
\end{eqnarray*}

Next we quantify $\| \overline{g} - g_0 \|_N$ . Using Lemma A.8 in \cite{Wang2007}, we have
\begin{eqnarray*}
C_N \equiv \sup_{f_1, f_2 \in \calG_N} \left| \frac{\langle f_1,f_2\rangle_N - \langle f_1,f_2\rangle}{\|f_1\| \|f_2\|} \right| = O \left( \frac{\log N}{(N h_N)^{1/2}} \right), \ \ a.s.
\end{eqnarray*}
Therefore, noting $\overline{g}, \wt{g} \in \calG_N$, we have
$$\| \overline{g} - g_0 \|_N \leq  \| \overline{g} - \wt{g} \|_N + \| \wt{g} - g_0 \|_N= O_P \left( J_N^{1/2}  N^{-1/2} + h_N^p\right). $$
\hfill $\Box$

\bigskip

\noindent{\it Proof of Corollary \ref{corollary1}}. For homogenous massive data, $\wh{\bbeta}^{(j)}$, $j=1,\ldots,s$, are i.i.d.~random vectors. \cite{li2013} showed that if $E(\wh{\bbeta}^{(j)} - \bbeta_0 )= o(N^{-1/2})$, then $\overline{\bbeta}$ is as efficient as $\wh{\bbeta}$, which is defined as via the following minimization using all $N$ observations,
$$ (\wh{\bbeta}, \wh{g}) =\argmin_{\stgbeta \in {\mathcal R}^d,\ g \in \calG_N} \frac{1}{2} \sum_{j=1}^s \sum_{i \in G_j} \left[ Y_i - \boldX_i\trans \bbeta - g(\boldZ_i) \right]^2.$$
\cite{liu2011} showed that
$$  \sqrt{N} (\wh{\bbeta} - \bbeta_0 ) \xrightarrow{d} \calN \big( \bfzero, \sigma^2 \boldD^{-1} \big). $$
Therefore it suffices to show that $E(\wh{\bbeta}^{(j)} - \bbeta_0 )= o(N^{-1/2})$. Following the proof of Theorem \ref{theorem1}, we have
\begin{eqnarray*}
&& \left( \frac{1}{n} \sum_{i \in G_j} \wt{\boldX}_i^{\otimes 2} + \frac{1}{n} \sum_{i \in G_j} \wt{\boldX}_i \Gamma(\boldZ_i)\trans \right) (\wh{\bbeta}^{(j)} - \bbeta_0 ) \\
&=& \frac{1}{n}\sum_{i \in G_j} \varepsilon_i \wt{\boldX}_i - \frac{1}{n} \sum_{i \in G_j} \left( \wh{g}^{(j)}(\boldZ_i) - g_0(\boldZ_i) \right) \wt{\boldX}_i,
\end{eqnarray*}
and it follows
\begin{eqnarray*}
\wh{\bbeta}^{(j)} - \bbeta_0 &=& \left( \frac{1}{n} \sum_{i \in G_j} \wt{\boldX}_i^{\otimes 2} + \frac{1}{n} \sum_{i \in G_j} \wt{\boldX}_i \Gamma(\boldZ_i)\trans \right)^{-1} \times \\
&& \left[\frac{1}{n}\sum_{i \in G_j} \varepsilon_i \wt{\boldX}_i - \frac{1}{n} \sum_{i \in G_j} \left( \wh{g}^{(j)}(\boldZ_i) - g_0(\boldZ_i) \right) \wt{\boldX}_i \right].
\end{eqnarray*}
Under Assumption A3 and the fact that $E(\phi(\boldZ) \wt{\boldX})=0$ for any measurable function $\phi$, we can show that
$$ 0 < c \leq E \left\| \frac{1}{n} \sum_{i \in G_j} \wt{\boldX}_i^{\otimes 2} + \frac{1}{n} \sum_{i \in G_j} \wt{\boldX}_i \Gamma(\boldZ_i)\trans \right\|_2 \leq C,$$
where $c$ and $C$ are some positive constants. Moreover, we have
\begin{eqnarray*}
E \left\{ \frac{1}{n} \sum_{i \in G_j} \left( \wh{g}^{(j)}(\boldZ_i) - g_0(\boldZ_i) \right) \wt{\boldX}_i \right\}= E \left( \wh{g}^{(j)}(\boldZ) - g_0(\boldZ) \right) \wt{\boldX}_i= O(h_N^p)=o(N^{-1/2}).
\end{eqnarray*}
Therefore, if $n\gg N^{1/2}$, by Cauchy-Schwarz inequality we have $E(\wh{\bbeta}^{(j)} - \bbeta_0 )=o(N^{-1/2})$. $\Box$

\bigskip

\noindent{\it Proof of Theorem \ref{theorem3}}.
The estimating equation is
\begin{eqnarray*}
\sum_{i \in G_j} \boldX_i \left\{Y_i - \boldX_i\trans \breve{\bbeta}^{(j)} - \overline{g}(\boldZ_i)\right\}=\bfzero.
\end{eqnarray*}
Since $Y_i = \boldX_i\trans \bbeta_0^{(j)} + g_0(\boldZ_i) + \varepsilon_i$, we have
\begin{eqnarray*}
\sqrt{n} \left( \breve{\bbeta}^{(j)} - \bbeta_0^{(j)} \right) =n^{-1/2} \sum_{i \in G_j} \left( \boldA_n^{(j)} \right)^{-1} \boldX_i\trans \varepsilon_i + n^{-1/2} \sum_{i \in G_j} \left( \boldA_n^{(j)} \right)^{-1} \boldX_i (g_0(\boldZ_i) - \overline{g}(\boldZ_i)).
\end{eqnarray*}
Considering the first term on the right hand side of the above equation, we have
\begin{eqnarray*}
n^{-1/2} \sum_{i \in G_j} \left( \boldA_n^{(j)} \right)^{-1} \boldX_i\trans \varepsilon_i \xrightarrow{d} \calN(\bfzero, \sigma^2\boldA^{-1}).
\end{eqnarray*}
Consider the second term on the right hand side. Let $w_2(\boldZ,g)=g(\boldZ)\left( \boldA_n^{(j)} \right)^{-1} \boldX$. We have
$$ E\big\| w_2(\boldZ,\overline{g}) - w_2(\boldZ,g_0) \big\|^2 =E \left\| (\overline{g}(\boldZ_i)-g_0(\boldZ_i)) \left( \boldA_n^{(j)} \right)^{-1} \boldX_i \right\|^2 \leq O\big(E \| \overline{g} -g_0 \|_2^2 \big). $$
By Lemma A.2 of \cite{Huang1999}, the logarithm of the $\varepsilon$-bracketing number of the class of functions $\calA_2(\delta)=\{ w_2(\cdot, \overline{g}) - s(\cdot,g_0) : \overline{g} \in \calG_N, \| \overline{g}-g_0 \|_2 \leq \delta \}$ is $c\{ (J_N - \varrho) \log (\delta/\varepsilon) + \log(\delta^{-1}) \}$. Thus, the corresponding entropy integral $J_{[\ ]}(\delta, \calA_2(\delta), \| \cdot \|_2 ) \leq c \delta \{ (J_N - \varrho)^{1/2} + \log^{1/2} (\delta^{-1}) \}$. According to Lemma 7 of \cite{Stone1986} and Theorem \ref{theorem2}, $\| \overline{g} - g_0 \|_{\infty} \leq c J_N^{1/2}\| \overline{g} - g_0 \|_{2}=O_P(J_N N^{-1/2} + J_N^{1/2} h_N^p)$. Let $r_{N}^{-1}=J_N^{1/2} N^{-1/2} +h_N^p$, then
\begin{eqnarray*}
&& E\left| \frac{1}{n} \sum_{i \in G_j} \left\{(\overline{g}(\boldZ_i)-g_0(\boldZ_i)) \left( \boldA_n^{(j)} \right)^{-1} \boldX_i \right\}\right. \\
&&\left.\hspace{1cm} - E \left\{ (\overline{g}(\boldZ_i)-g_0(\boldZ_i)) \left( \boldA_n^{(j)} \right)^{-1} \boldX_i \right\} \right| \\
&\leq& n^{-1/2} C r_{N}^{-1} \left\{ (J_N +\varrho)^{1/2} + \log^{1/2}(r_{N})\right\} \\
&&\hspace{1cm} \times
\left[ 1 + \frac{c r_{N}^{-1} \left\{ (J_N+ \varrho)^{1/2} + \log^{1/2}(r_{N})\right\}}{r_{N}^{-2}\sqrt{n}} C_0 \right]\\
&\leq& O(s^{1/2}) n^{-1/2} C r_{N}^{-1} \left\{ (J_N +\varrho)^{1/2} + \log^{1/2}(r_{N})\right\} \\
&=& O(n^{-1/2}) \times O\left( J_N n^{-1/2} + s^{1/2}J_N^{1/2} h_N^p \right) \\
&=& O(n^{-1/2}) \times O\left( J_N n^{-1/2} + \big( n^{-1} N^{1+ q(1-2p)} \big)^{1/2} \right) \\
&\leq& O(n^{-1/2}) \times O\left( J_N n^{-1/2} + \big( n^{-1} N^{1/(2p)} \big)^{1/2} \right),
\end{eqnarray*}
where the last inequality is due to the condition that $J_N \gg N^{1/(2p)}$.
The condition that $J_N^2 \ll n$ implies that $O( J_N n^{-1/2})=o(1)$ and $n \gg N^{1/(2p)}$ to make sure that the above expectation has an order $o(n^{-1/2})$.
Furthermore,
$$ E \left\{ (\overline{g}(\boldZ_i)-g_0(\boldZ_i)) \left( \boldA_n^{(j)} \right)^{-1} \boldX_i \right\}  \leq O(E\| \overline{g} - g_0 \|_{\infty}) =O(J_N  N^{-1/2}).  $$
Thus,
$$ n^{-1/2} \sum_{i \in G_j} \left( \boldA_n^{(j)} \right)^{-1} \boldX_i (g_0(\boldZ_i) - \overline{g}(\boldZ_i)) = O(J_N s^{-1/2}) + o_P(1) = o_P(1),$$
where the last equality is due to the condition that $J_N^2 \ll s$. Therefore, the theorem is proved. \hfill $\Box$

\bigskip

\noindent{\it Proof of Theorem \ref{theorem4}}. Under the null hypothesis, we have
$$ \sqrt{n} \boldQ \big( \wh\bbeta^{(j_1)} - \wh\bbeta^{(j_2)}\big)  =
\sqrt{n} \boldQ \big( \wh\bbeta^{(j_1)} -\bbeta_0^{(j_1)}\big) - \sqrt{n} \boldQ \big(\wh\bbeta^{(j_2)}-\bbeta_0^{(j_2)}\big).$$
By Theorem \ref{theorem1}, we have
$$\sqrt{n} \boldQ \big( \wh\bbeta^{(j_t)} -\bbeta_0^{(j_t)}\big) \xrightarrow{d} \calN \big( \bfzero, \sigma^2 \boldQ\boldD^{-1}\boldQ\trans \big),$$
where $t=1$ or $2$. Therefore, $ \sqrt{n} \boldQ \big( \wh\bbeta^{(j_1)} - \wh\bbeta^{(j_2)}\big) \xrightarrow{d} \calN \big( \bfzero, 2\sigma^2 \boldQ \boldD^{-1} \boldQ\trans \big).$

Consider the second result. According to the proof of Theorem \ref{theorem3}, we have
\begin{eqnarray*}
&& \sqrt{n} \left( \breve{\bbeta}^{(j)} - \bbeta_0^{(j)} \right) =n^{-1/2} \sum_{i \in G_j} \left( \boldA_n^{(j)} \right)^{-1} \boldX_i\trans \varepsilon_i \\
&&\hspace{2cm} + n^{-1/2} \sum_{i \in G_j} \left( \boldA_n^{(j)} \right)^{-1} \boldX_i (g_0(\boldZ_i) - \overline{g}(\boldZ_i)).
\end{eqnarray*}
Thus, with similar arguments in the proof of Theorem \ref{theorem3}, we have
\begin{eqnarray*}
&& \sqrt{n} \boldQ \big( \breve\bbeta^{(j_1)} - \breve\bbeta^{(j_2)}\big) =  \sqrt{n} \boldQ \big( \breve\bbeta^{(j_1)} -\bbeta_0^{(j_1)}\big) - \sqrt{n} \boldQ \big(\breve\bbeta^{(j_2)}-\bbeta_0^{(j_2)}\big) \\
&=&  n^{-1/2} \sum_{i \in G_{j_1}} \boldQ \left( \boldA_n^{(j_1)} \right)^{-1} \boldX_i\trans \varepsilon_i + n^{-1/2} \sum_{i \in G_{j_1}} \boldQ\left( \boldA_n^{(j_1)} \right)^{-1} \boldX_i (g_0(\boldZ_i) - \overline{g}(\boldZ_i)) \\
&& -  n^{-1/2} \sum_{i \in G_{j_2}} \boldQ \left( \boldA_n^{(j_2)} \right)^{-1} \boldX_i\trans \varepsilon_i - n^{-1/2} \sum_{i \in G_{j_2}} \boldQ\left( \boldA_n^{(j_2)} \right)^{-1} \boldX_i (g_0(\boldZ_i) - \overline{g}(\boldZ_i)) \\
&=& n^{-1/2} \sum_{i \in G_{j_1}} \boldQ \left( \boldA_n^{(j_1)} \right)^{-1} \boldX_i\trans \varepsilon_i - n^{-1/2} \sum_{i \in G_{j_2}} \boldQ \left( \boldA_n^{(j_2)} \right)^{-1} \boldX_i\trans \varepsilon_i + o_p(1) \\
&\xrightarrow{d} & \calN \big( \bfzero, 2\sigma^2 \boldQ\boldA^{-1}\boldQ\trans \big).
\end{eqnarray*}
Therefore, the second result is also proved. $\Box$

\bigskip

\noindent{\it Proof of Theorem \ref{theorem5}}. First of all, based on the proof of Theorem \ref{theorem3}, we have
\begin{eqnarray*}
\sqrt{n} \left( \breve{\bbeta}^{(j)} - \check\bbeta_0^{(j)} \right) =n^{-1/2} \sum_{i \in G_j} \left( \wh\boldA_n^{(j)} \right)^{-1} \boldX_i\trans \varepsilon_i + R_i.
\end{eqnarray*}
where $R_i = n^{-1/2} \sum_{i \in G_j} \left( \wh\boldA_n^{(j)} \right)^{-1} \boldX_i (g_0(\boldZ_i) - \overline{g}(\boldZ_i))$.
Also, we have shown there
$$ \max_{j \in \calS} \| R_i\|_\infty = O_P(J_N s^{-\frac{1}{2}}).$$
Noting that $s \gg J_N^2 \log(pd)$, it follows
$$ \max_{j \in \calS} \| R_i\|_\infty = o_P(\log^{-\frac{1}{2}}(pd)).$$
Then, the rest of the proof follows a similar line with the counterpart of Theorem 3.9 in \cite{Zhao2016} by discussing the asymptotic differences between approximation terms. \hfill $\Box$

\bigskip

\noindent{\it Proof of Theorem \ref{theorem6}}. The estimates for $g_k$ and $g_{-k}$ depends on B-spline basis expansions. Denote $\bgamma_k$ as coefficients for basis $\boldB_k(Z_k)$ of the $k$th function $g_k$, and $\bgamma_{-k}$ as coefficients for basis $\boldB_{-k}(\boldZ_{-k})$ of $g_{-k}$. With a little abuse of notations, we write
\begin{eqnarray*}
&& L_{nk}^{(j)}(\bbeta,\bgamma_k,\bgamma_{-k}) \\
&=& \frac{1}{2n} \sum_{i \in G_j} \left[ Y_i - \boldX_i\trans \bbeta - \boldB_k\trans(Z_{ik}) \bgamma_k - \boldB_{-k}\trans(\boldZ_{i,-k}) \bgamma_{-k} \right]^2 \\
&=& \frac{1}{2n} \sum_{i \in G_j} \Big[ \varepsilon_i + \boldX_i\trans (\bbeta_0^{(j)} - \bbeta) + \boldB_k\trans(Z_{ik}) (\wt{\bgamma}_k^{(j)} - \bgamma_k) \\
&& + (g_{0k}^{(j)}(Z_{ik}) - \wt{g}_k^{(j)}(Z_{ik})) + (g_{0,-k}(\boldZ_{i,-k}) -\boldB_{-k}\trans(\boldZ_{i,-k})\bgamma_{-k}) \Big],
\end{eqnarray*}
where $\wt{g}_k^{(j)}(Z_{ik}) = \boldB_k\trans(Z_{ik}) \bgamma_k^{(j)}$.

Consider the derivative of $L_{nk}^{(j)}(\bbeta,\bgamma_k,\bgamma_{-k})$ with respect to $\bgamma_k$:
\begin{eqnarray*}
\nabla L_{nk}^{(j)}(\bbeta,\bgamma_k,\bgamma_{-k}) &=& -\frac{1}{n} \sum_{i \in G_j} \varepsilon_i \boldB_k(Z_{ik}) + \frac{1}{n} \sum_{i \in G_j} \boldB_k(Z_{ik})^{\otimes 2}(\bgamma_k - \wt{\bgamma}_k^{(j)} ) \\
&& + \frac{1}{n} \sum_{i \in G_j} (\wt{g}_k^{(j)}(Z_{ik}) - g_{0k}^{(j)}(Z_{ik}) )\boldB_k(Z_{ik}) \\
&& + \frac{1}{n} \sum_{i \in G_j}  (\boldB_{-k}\trans(\boldZ_{i,-k})\bgamma_{-k} - g_{0,-k}(\boldZ_{i,-k}))\boldB_k(Z_{ik}) \\
&& + \frac{1}{n} \sum_{i \in G_j} \boldX_i\trans ( \bbeta - \bbeta_0^{(j)}) \boldB_k(Z_{ik}).
\end{eqnarray*}

By plugging estimates based on the $j$th sub-population, it follows that
\begin{eqnarray*}
\nabla L_{nk}^{(j)}(\wh\bbeta^{(j)},\wh\bgamma_k^{(j)},\wh\bgamma_{-k}^{(j)}) &=& -\frac{1}{n} \sum_{i \in G_j} \varepsilon_i \boldB_k(Z_{ik}) + \frac{1}{n} \sum_{i \in G_j} \boldB_k(Z_{ik})^{\otimes 2}(\wh\bgamma_k^{(j)} - \wt{\bgamma}_k^{(j)} ) \\
&& + \frac{1}{n} \sum_{i \in G_j} (\wt{g}_k^{(j)}(Z_{ik}) - g_{0k}^{(j)}(Z_{ik}) )\boldB_k(Z_{ik}) \\
&& + \frac{1}{n} \sum_{i \in G_j}  (\boldB_{-k}\trans(\boldZ_{i,-k})\wh\bgamma_{-k}^{(j)} - g_{0,-k}(\boldZ_{i,-k}))\boldB_k(Z_{ik}) \\
&& + \frac{1}{n} \sum_{i \in G_j} \boldX_i\trans ( \wh\bbeta^{(j)} - \bbeta_0^{(j)}) \boldB_k(Z_{ik}), \\
\nabla L_{nk}^{(j)}(\wh\bbeta^{(j)},\wt\bgamma_k^{(j)},\wh\bgamma_{-k}^{(j)}) &=& -\frac{1}{n} \sum_{i \in G_j} \varepsilon_i \boldB_k(Z_{ik}) \\
&& + \frac{1}{n} \sum_{i \in G_j} (\wt{g}_k^{(j)}(Z_{ik}) - g_{0k}^{(j)}(Z_{ik}) )\boldB_k(Z_{ik}) \\
&& + \frac{1}{n} \sum_{i \in G_j}  (\boldB_{-k}\trans(\boldZ_{i,-k})\wh\bgamma_{-k}^{(j)} - g_{0,-k}(\boldZ_{i,-k}))\boldB_k(Z_{ik}) \\
&& + \frac{1}{n} \sum_{i \in G_j} \boldX_i\trans ( \wh\bbeta^{(j)} - \bbeta_0^{(j)}) \boldB_k(Z_{ik}).
\end{eqnarray*}

Then, we obtain
\begin{eqnarray*}
\nabla L_{nk}^{(j)}(\wh\bbeta^{(j)},\wh\bgamma_k^{(j)},\wh\bgamma_{-k}^{(j)}) = \nabla L_{nk}^{(j)}(\wh\bbeta^{(j)},\wt\bgamma_k^{(j)},\wh\bgamma_{-k}^{(j)}) + \frac{1}{n} \sum_{i \in G_j} \boldB_k(Z_{ik})^{\otimes 2}(\wh\bgamma_k^{(j)} - \wt{\bgamma}_k^{(j)} ).
\end{eqnarray*}
Due to the stationary equation to \eqref{eq:j-popu-equi}, we have
\begin{eqnarray*}
\nabla L_{nk}^{(j)}(\wh\bbeta^{(j)},\wt\bgamma_k^{(j)},\wh\bgamma_{-k}^{(j)}) = - \frac{1}{n} \sum_{i \in G_j} \boldB_k(Z_{ik})^{\otimes 2}(\wh\bgamma_k^{(j)} - \wt{\bgamma}_k^{(j)} ).
\end{eqnarray*}

Let $E \{\boldB_k(Z_{ik})^{\otimes 2}\} = \bSigma_k$. According to Lemma \ref{lemmaA2}, we know $\bSigma_k$'s eigenvalues are bounded and away from 0. Thus,
\begin{eqnarray*}
&&\nabla L_{nk}^{(j)}(\wh\bbeta^{(j)},\wh\bgamma_k^{(j)},\wh\bgamma_{-k}^{(j)}) - \nabla L_{nk}^{(j+1)}(\wh\bbeta^{(j+1)},\wh\bgamma_k^{(j+1)},\wh\bgamma_{-k}^{(j+1)}) \\
&=& - (\bSigma_k + o_P(1))(\wh\bgamma_k^{(j)} - \wh\bgamma_k^{(j+1)}),
\end{eqnarray*}
where the last equation is due to $\wt{\bgamma}_k^{(j)}=\wt{\bgamma}_k^{(j+1)}$ under the null hypothesis.

By Taylor expansion, we have
\begin{eqnarray*}
&& \nabla L_{nk}^{(j)}(\wh\bbeta^{(j)},\wh\bgamma_k^{(j+1)},\wh\bgamma_{-k}^{(j)}) \\
&=& \nabla L_{nk}^{(j)}(\wh\bbeta^{(j)},\wt\bgamma_k^{(j+1)},\wh\bgamma_{-k}^{(j)}) + \frac{1}{n} \sum_{i \in G_j} \boldB_k(Z_{ik})^{\otimes 2}(\wh\bgamma_k^{(j+1)} - \wt{\bgamma}_k^{(j+1)} ) \\
&=& \nabla L_{nk}^{(j)}(\wh\bbeta^{(j)},\wt\bgamma_k^{(j)},\wh\bgamma_{-k}^{(j)}) - \nabla L_{nk}^{(j+1)}(\wh\bbeta^{(j+1)},\wt\bgamma_k^{(j+1)},\wh\bgamma_{-k}^{(j+1)}) + o_P(1).
\end{eqnarray*}
Then, applying Taylor expansion again leads to
\begin{eqnarray*}
&& L_{nk}^{(j)}(\wh\bbeta^{(j)},\wh\bgamma_k^{(j)},\wh\bgamma_{-k}^{(j)}) - L_{nk}^{(j)}(\wh\bbeta^{(j)},\wh\bgamma_k^{(j+1)},\wh\bgamma_{-k}^{(j)}) \\
&=& \nabla L_{nk}^{(j)}(\wh\bbeta^{(j)},\wh\bgamma_k^{(j+1)},\wh\bgamma_{-k}^{(j)})\trans (\wh\bgamma_k^{(j)} - \wh\bgamma_k^{(j+1)}) \\
&& \quad + \frac{1}{2} (\wh\bgamma_k^{(j)} - \wh\bgamma_k^{(j+1)})\trans (\bSigma_k + o_P(1))(\wh\bgamma_k^{(j)} - \wh\bgamma_k^{(j+1)}) \\
&=& - \frac{1}{2} (\nabla L_{nk}^{(j)}(\wh\bbeta^{(j)},\wt\bgamma_k^{(j)},\wh\bgamma_{-k}^{(j)}) - \nabla L_{nk}^{(j+1)}(\wh\bbeta^{(j+1)},\wt\bgamma_k^{(j+1)},\wh\bgamma_{-k}^{(j+1)}))\trans (\bSigma_k + o_P(1)) \\
&& \ \ \times (\nabla L_{nk}^{(j)}(\wh\bbeta^{(j)},\wt\bgamma_k^{(j)},\wh\bgamma_{-k}^{(j)}) - \nabla L_{nk}^{(j+1)}(\wh\bbeta^{(j+1)},\wt\bgamma_k^{(j+1)},\wh\bgamma_{-k}^{(j+1)})).
\end{eqnarray*}

Based on Lemmas \ref{lemmaA1}-\ref{lemmaA5}, we can verify that
$$\nabla L_{nk}^{(j)}(\wh\bbeta^{(j)},\wt\bgamma_k^{(j)},\wh\bgamma_{-k}^{(j)}) = -\frac{1}{n} \sum_{i \in G_j} \varepsilon_i \boldB_k(Z_{ik}) + R_N,$$
where $\|R_N \|_2=O_P((J_N^{1/2}n^{-1/2} + h_N^p)J_N^{1/2}).$
It is straightforward to have:
\begin{eqnarray*}\nabla L_{nk}^{(j)}(\wh\bbeta^{(j)},\wt\bgamma_k^{(j)},\wh\bgamma_{-k}^{(j)}) - \nabla L_{nk}^{(j+1)}(\wh\bbeta^{(j+1)},\wt\bgamma_k^{(j+1)},\wh\bgamma_{-k}^{(j+1)}) \\ \stackrel{d}{=} -\frac{1}{n} \sum_{i \in G_j \bigcup G_{j+1}} \varepsilon_i \boldB_k(Z_{ik}) + R_N.
\end{eqnarray*}

Therefore,
\begin{eqnarray*}
&& n{\rm LRT}_{nk}^s\\
&=& -n\sum_{j=1}^{s-1} \Bigg\{ \left[\frac{1}{n} \sum_{i \in G_j \bigcup G_{j+1}} \varepsilon_i \boldB_k(Z_{ik}) \right]\trans \bSigma_k^{-1} \left[\frac{1}{n} \sum_{i \in G_j \bigcup G_{j+1}} \varepsilon_i \boldB_k(Z_{ik}) \right] \\
&& \quad + O_P((J_N^{1/2}n^{-1/2} + h_N^p)J_N^{1/2}n^{-1/2}) \Bigg\} \\
&=& -\sum_{j=1}^{s-1} \frac{1}{n} \sum_{i \in G_j \bigcup G_{j+1}} \varepsilon_i^2 \boldB_k(Z_{ik})\trans\bSigma_k^{-1} \boldB_k(Z_{ik}) \\
&& \quad - \sum_{j=1}^{s-1} \frac{1}{n} \sum_{(i_1 < i_2) \in G_j \bigcup G_{j+1}} 2 \varepsilon_{i_1}\varepsilon_{i_2} \boldB_k(Z_{{i_1}k})\trans\bSigma_k^{-1} \boldB_k(Z_{i_2k}) \\
&& \quad \ \ \ + O_P((J_N^{1/2}n^{-1/2} + h_N^p)J_N n^{1/2}).
\end{eqnarray*}

The rate of the first term for $n{\rm LRT}_{nk}^s$ can be found investigating the moment
\begin{eqnarray*}
&& \frac{1}{n^2} \sum_{i \in G_j \bigcup G_{j+1}} E\left\{  \varepsilon_i^2 \boldB_k(Z_{ik})\trans\bSigma_k^{-1} \boldB_k(Z_{ik})  -E \varepsilon_i^2 \boldB_k(Z_{ik})\trans\bSigma_k^{-1} \boldB_k(Z_{ik})\right\}^2 \\
&\leq& \frac{2}{n} E \varepsilon_i^4 (\boldB_k(Z_{ik})\trans\bSigma_k^{-1} \boldB_k(Z_{ik}))^2  \\
&\leq& \frac{2C}{n}J_N^2 <\infty,
\end{eqnarray*}
where $C$ is some constant. Also, $E[\boldB_k(Z_{ik})\trans\bSigma_k^{-1} \boldB_k(Z_{ik})]=J_N$. Then, it follows
\begin{eqnarray}
\sum_{j=1}^{s-1} \frac{1}{n} \sum_{i \in G_j \bigcup G_{j+1}} \varepsilon_i^2 \boldB_k(Z_{ik})\trans\bSigma_k^{-1} \boldB_k(Z_{ik}) = 2 (s-1) \sigma^2 J_N + o_P(sJ_N). \label{eq:U_mean}
\end{eqnarray}

Let $W_{i_1 i_2}^{(j)} = 2 \varepsilon_{i_1}\varepsilon_{i_2} \boldB_k(Z_{{i_1}k})\trans\bSigma_k^{-1} \boldB_k(Z_{i_2k})$, and then define $$W(N)=\sum_{j=1}^{s-1}\sum_{(i_1 < i_2) \in G_j \bigcup G_{j+1}} W_{i_1 i_2}^{(j)}.$$
The next step is to show the asymptotic normality of $n^{-1}W(N)$, which follows a similar line with the proof of Theorem 4.1 in \cite{lu2016} using Proposition 3.2 in \cite{deJong1987}. First, it is easy to see $E(W_{i_1 i_2}^{(j)}|Z_{i_1 k})=0,\forall i_1 < i_2$. Second, bounds for several moments of $W(N)$ are derived. Define $\sigma(N)^2 = \textrm{Var}(W(N))$ and following quantities:
\begin{eqnarray*}
U_{I} &=& \sum_{j=1}^{s-1} \sum_{\substack{(i_1 < i_2) \\ \in G_j \bigcup G_{j+1}} } E[(W_{i_1 i_2}^{(j)})^4], \\
U_{II} &=& \sum_{j=1}^{s-1} \sum_{\substack{(i_1 < i_2 < i_3) \\ \in G_j \bigcup G_{j+1}}} \Big\{ E[(W_{i_1 i_2}^{(j)})^2 (W_{i_1 i_3}^{(j)})^2] + E[(W_{i_2 i_1}^{(j)})^2 (W_{i_2 i_3}^{(j)})^2] \\
&& \quad+ E[(W_{i_3 i_1}^{(j)})^2 (W_{i_3 i_2}^{(j)})^2] \Big\} \\
U_{III} &=& \sum_{j=1}^{s-1} \sum_{\substack{(i_1 < i_2 < i_3 < i_4) \\ \in G_j \bigcup G_{j+1}}} \Big\{
E[W_{i_1 i_2}^{(j)} W_{i_1 i_3}^{(j)}W_{i_4 i_2}^{(j)}W_{i_4 i_3}^{(j)}] +
E[W_{i_1 i_2}^{(j)} W_{i_1 i_4}^{(j)}W_{i_3 i_2}^{(j)}W_{i_3 i_4}^{(j)}] \\
&& \quad \quad + E[W_{i_1 i_3}^{(j)} W_{i_1 i_4}^{(j)}W_{i_2 i_3}^{(j)}W_{i_2 i_4}^{(j)}].
\Big\}
\end{eqnarray*}
By the definition of $\bSigma_k$ and $E(\varepsilon_i^4) < \infty$, we have
\begin{eqnarray*}
E[(W_{i_1 i_2}^{(j)})^4] &\leq& 2^4 E\left[\varepsilon_{i_1}^4\varepsilon_{i_2}^4 (\boldB_k(Z_{{i_1}k})\trans\bSigma_k^{-1} \boldB_k(Z_{i_2k}))^4\right] =O(J_N^4) \\
E[(W_{i_1 i_2}^{(j)})^2 (W_{i_1 i_3}^{(j)})^2] &\leq& E[(W_{i_1 i_2}^{(j)})^4]  =O(J_N^4) \\
E[W_{i_1 i_2}^{(j)} W_{i_1 i_3}^{(j)}W_{i_4 i_2}^{(j)}W_{i_4 i_3}^{(j)}] &=& O(J_N)
\end{eqnarray*}
Then it follows that
$$U_{I} = O(s n^2 J_N^4), \ U_{II}=O(s n^3 J_N^4), \ U_{III}=O(s n^4 J_N^4).$$

Third, $\sigma(N)^4 \sim \left[ (s-1)\binom{2n}{2}4\sigma^4 J_N+ 2(s-2)\binom{n}{2}4\sigma^4 J_N\right]^2\sim (12 \sigma^4 (s-1) n^2 J_N)^2$ because $E[(W_{i_1 i_2}^{(j)})^2]=4\sigma^4 J_N$. Since $U_{I}, U_{II}$ and $U_{III}$ have a smaller order than $\sigma(N)^4$ under the condition $J_N^2 \ll n$, Proposition 3.2 in \cite{deJong1987} shows that
\begin{eqnarray} \label{eq:U_stat}
\frac{1}{\sqrt{12 \sigma^4 (s-1) n^2 J_N}} W(N)  \xrightarrow{d} \calN(0,1).
\end{eqnarray}

Last, under the condition $J_n \gg n^{\frac{1}{2p}}$ from Theorem \ref{theorem1} and $s^2 \gg J_N$, we know $(J_N^{1/2}n^{-1/2} + h_N^p)J_N n^{1/2} \ll (J_N^{1/2}+ n^{-1/2} n^{1/2} )J_N \ll s J_N$. Therefore, by combining \eqref{eq:U_mean} and \eqref{eq:U_stat}, we have
\begin{eqnarray*}
\frac{1}{\sqrt{2 \frac{2}{3}(s-1) J_N}} \left(-\frac{1}{3\sigma^2}n{\rm LRT}_{nk}^s - \frac{2}{3}(s-1) J_N\right)  \xrightarrow{d} \calN(0,1).
\end{eqnarray*}
\hfill $\Box$

%

\bibliographystyle{imsart-nameyear}
\bibliography{refs_APLM}

\end{document}